\documentclass[conference]{IEEEtran}

\usepackage{cite}
\usepackage{amsmath,amssymb,amsfonts}
\usepackage{algorithmic}
\usepackage{graphicx}
\usepackage{textcomp}
\usepackage{xcolor}
\usepackage{subcaption}
\usepackage{enumitem}

\usepackage{booktabs} 
\usepackage{url}
\usepackage{wrapfig}
\usepackage{hyperref}

\def\BibTeX{{\rm B\kern-.05em{\sc i\kern-.025em b}\kern-.08em
    T\kern-.1667em\lower.7ex\hbox{E}\kern-.125emX}}

\begin{document}

\title{GREEN-CODE: Learning to Optimize Energy Efficiency in LLM-based Code Generation}


\author{

    \IEEEauthorblockN{Shashikant Ilager}
    \IEEEauthorblockA{
      University of Amsterdam, The Netherlands 
    }
       \and
      \IEEEauthorblockN{Lukas Florian Briem}
    \IEEEauthorblockA{
        TU Wien, Austria
    }
    \and
    \IEEEauthorblockN{Ivona Brandic}
    \IEEEauthorblockA{
        TU Wien, Austria 
    }
}

\IEEEaftertitletext{\vspace{-5\baselineskip}}

\maketitle

\begin{abstract} 

Large Language Models (LLMs) are becoming integral to daily life, showcasing their vast potential across various Natural Language Processing (NLP) tasks. Beyond NLP, LLMs are increasingly used in software development tasks,  such as code completion, modification, bug fixing, and code translation. Software engineers widely use tools like GitHub Copilot and Amazon Q, streamlining workflows and automating tasks with high accuracy. While the resource and energy intensity of LLM training is often highlighted, inference can be even more resource-intensive over time, as it's a continuous process with a high number of invocations. Therefore, developing resource-efficient alternatives for LLM inference is crucial for sustainability. This work proposes GREEN-CODE, a framework for energy-aware code generation in LLMs. GREEN-CODE performs dynamic early exit during LLM inference. We train a  Reinforcement Learning (RL) agent that learns to balance the trade-offs between accuracy, latency, and energy consumption. Our approach is evaluated on two open-source LLMs, Llama 3.2 3B and OPT 2.7B, using the JavaCorpus and PY150 datasets. Results show that our method reduces the energy consumption between 23--50\% on average for code generation tasks without significantly affecting accuracy.
\end{abstract}

\begin{IEEEkeywords}
Large Language Models (LLMs), Green AI, Sustainable AI, Code Generation, Pruning, Early Exits
\end{IEEEkeywords}

\section{Introduction}

Large Language Models (LLMs) are becoming pervasive in software development, aiding in tasks such as code completion, modification, bug fixing, and code translation. Tools like GitHub Copilot~\cite{copilot} and Amazon Q Developer~\cite{amazonq} have already been adopted by software engineers, automating their tasks with high accuracy and boosting productivity~\cite{yetiştiren2023}. However, a significant issue with these LLM-based code generation tools is their high resource consumption, leading to substantial energy and carbon footprints.

While the resource-intensive nature of LLM training is often highlighted, the energy demands of inference—the process of using the models in a deployed setting are less frequently discussed. \textit{However, over time, inference often becomes more resource-hungry than training as pre-training is done only once while inference runs continuously, scaling for millions of users.} For example, a study~\cite{deVries2023} estimates that ChatGPT's daily energy consumption during inference is about 564 MWh, meaning it uses the same amount of energy in just 2.3 days as it did for the entire pretraining process of GPT-3's, which consumed 1,287 MWh.

Tools that rely on code generation models are continuously used during software development, e.g., suggesting code completions after every change in a file. Therefore, efficient methods tailored to this domain are urgently needed. Moreover, code generation tasks are often real-time workloads, requiring faster and more accurate predictions to improve the Quality of Service (QoS) or developer experience. In some cases, LLMs might need to run on local devices to preserve the privacy of the codebase. Such deployments lack high-performance large-scale hardware, making energy-efficient and latency-sensitive code generation even crucial.

There exist many optimization methods for LLM inference,  including quantization ~\cite{Dettmers2024,Chee2024,Kuzmin2022,Liu2021}, knowledge distillation ~\cite{jiao2020,  li2024, timiryasov2023}, and pruning ~\cite{ma2023, yang2024, wei2024}. However, these techniques can result in significantly reduced accuracy or introduce an irreversible accuracy loss in the optimization process. In contrast, \textbf{early exiting }offers a promising approach for optimizing runtime computational and energy costs based on application requirements and resource constraints ~\cite{ xin2020, xin2021, Schuster2022, Sun2024}. The model exits at specific layers of the transformer model by skipping deeper layers,  based on runtime decisions and contexts, reducing computation. Unlike other methods, early exiting can be done fully dynamically at runtime, allowing real-time decisions on when to exit early. This is crucial for code generation, where we can dynamically choose the required accuracy level based on resource and energy constraints.

Dynamic early exit for code generation presents several challenges.
\textit{First,} it is particularly challenging to determine the optimal exit points.
Existing approaches mainly use either confidence or entropy-based ("score-based") exit triggers~\cite{ Schuster2022, xin2020} or train a classifier~\cite{xin2021, Schuster2022, Sun2024} for exit decisions. Both approaches are static, as they do not adapt well to shifts in the problem domain or contextual changes, since reformulation/retraining must be done manually. \textit{Second}, current approaches use multiple Language Model Head (LM Head) architectures for early exit~\cite{Schuster2022, xin2020, Sun2024, Zeng2024}, introducing additional language model heads to enable early exits, where these heads learn from intermediate hidden states to predict tokens. However, these additional heads significantly increase the number of parameters that must be fully trained. \textit{Finally}, most of the existing studies do not consider energy consumption as a key metric for optimization, which is complex due to non-linear dependencies between exit layers, accuracy,  model and hardware architectures, and energy ~\cite{ilager_gpu_2024, Schuster2022, xin2020, xin2021, Sun2024}.

To address these limitations, in this paper, we propose GREEN-CODE, a learning-based framework for energy-efficient code generation through dynamic early exits. %
Initially, we introduce a fine-tuning method for code generation using an aggregated loss function, where intermediate layers can be made suitable for decoding~\cite{varshney2023} with a single LM Head. This eliminates the need for multiple LM Heads and their associated computational overhead. Next, we develop an adaptive Reinforcement Learning (RL)-based approach that dynamically adapts to the trade-offs between exit points, quality of code generation (accuracy), and energy consumption.

We implement the prototype system and evaluate the proposed techniques on our testbed. Furthermore, we integrated our solution into a VS Code extension as a demonstration of the practical use case of our framework, capable of leveraging the early exit strategy using a self-hosted fine-tuned model. We use with two popular LLMs, OPT and Llama 3.2, with 2.7B and 3B parameters, respectively. We evaluate two benchmark datasets, JavaCorpus and PY150.  GREEN-CODE prioritizes energy consumption as a primary efficiency metric to provide a more comprehensive assessment of model performance, alongside traditional performance metrics. Experiments conducted on the NVIDIA RTX 8000 shows that GREEN-CODE achieves comparable accuracy to the baseline using all layers of a non-fine-tuned model, while significantly reducing energy consumption between 23--50\%, on average. 
\\
In summary, we make the following \textbf{\textit{key contributions:}}
\\
\begin{itemize}
    \item  We \textbf{propose fine-tuning models for code generation tasks based on weighted aggregated loss}, enabling models to exit at arbitrary layers without introducing additional LM heads.
    \item  We design and \textbf{present GREEN-CODE framework,} an RL-based approach for energy efficient code generation through dynamic early exits. 
    \item  We \textbf{implement a prototype system} and evaluate using state-of-the-art LLMs and standard benchmark datasets.
    \item  We demonstrate the efficiency and efficacy of GREEN-CODE with \textbf{extensive experiments, comparing results with baseline} models across several performance and efficiency metrics.
\end{itemize}

\section{ Background Motivation}\label{sec:motivation}
GREEN-CODE's solution is based on the intuition that shallow layers can predict many tokens correctly, while deeper layer inferences are required for fewer token predictions. To understand the potential of early exiting, we conducted preliminary experiments with the 3B parameter model  Llama 3.2. The model is fine-tuned with aggregated loss (as introduced later in Section~\ref{sec:finetune}) that is capable of exiting at intermediate layers, generating predictions through a single LM head. We used fixed exiting in this experiment. 

\begin{figure}[!t]
\centering 
\includegraphics[width=\linewidth]{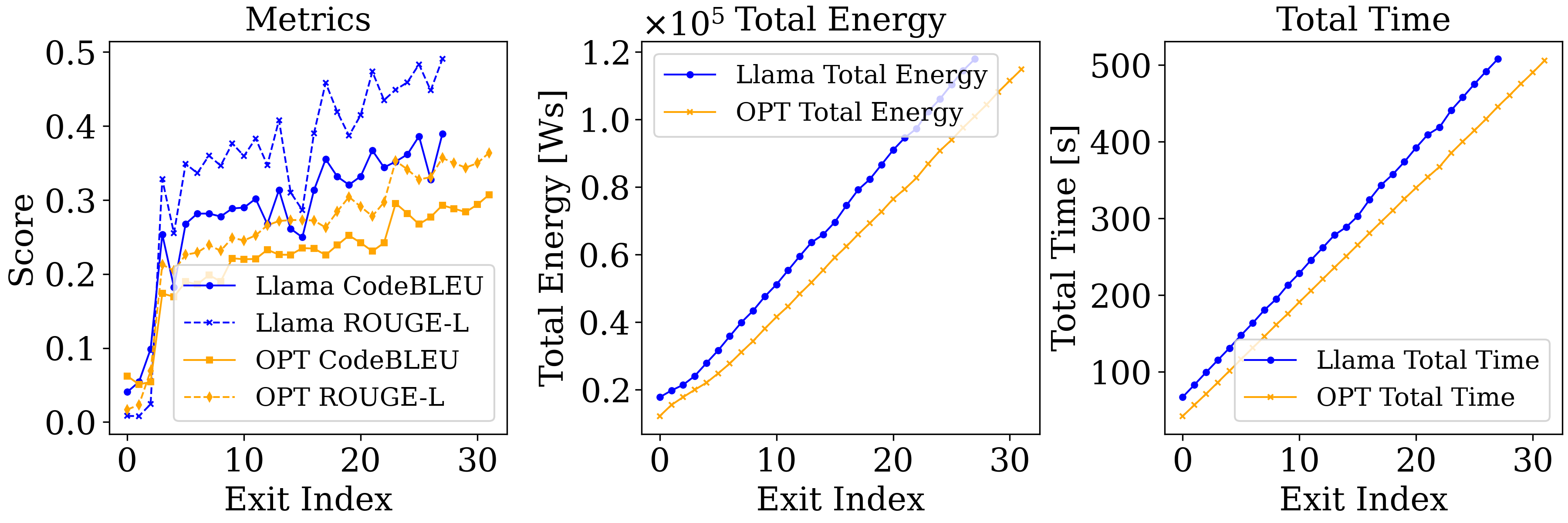}
\caption{Performance, energy  and latency of OPT-2.7B and Llama3.2-3B on JavaCorpus and PY150 with fixed exiting.} 
\label{fig:introduction_motivation}
\end{figure}

Figure~\ref{fig:introduction_motivation} shows the results (metrics including RougeL and CodeBLEU scores, energy consumption, and latency) from this experiment. As observed, with appropriate fine-tuning of the LLM, even shallow layers can predict many tokens correctly. For example, CodeBLEU on layer 10 of Llama is 0.3, while the final layer achieves 0.4, showing the potential of early exiting. In addition, energy and latency increase when the model utilizes deeper layers during inference. This provides an opportunity to develop a dynamic early exiting method by predicting the exit points for tokens during runtime. Moreover, early exiting offers a flexible trade-off between accuracy and efficiency, which other methods often cannot achieve, and it is applicable to most of the model architectures.

However, determining the right layer for each inference task is difficult since accuracy and exit layers do not scale linearly due to the complexities of pre-trained model architectures. Similarly, the trade-off between energy consumption, exit layer, and input data distribution modeling is also complex. Therefore, we need a dynamic and adaptive approach and a controller to make runtime decisions on exit layer strategies considering contextual information, model architectures, as well as energy and latency constraints.

\label{sec:background}
\section{ System Model and Methodology}
In this section, we  first present the system model of GREEN-CODE framework.  Then, describe the rationale behind datasets and LLMs selected. Next, we discuss the  LLM fine-tuning process designed to enable early exiting, providing an overview of our methodology.

\subsection{System Model}
\begin{figure}[!t]
\centering 
\includegraphics[width=\columnwidth]{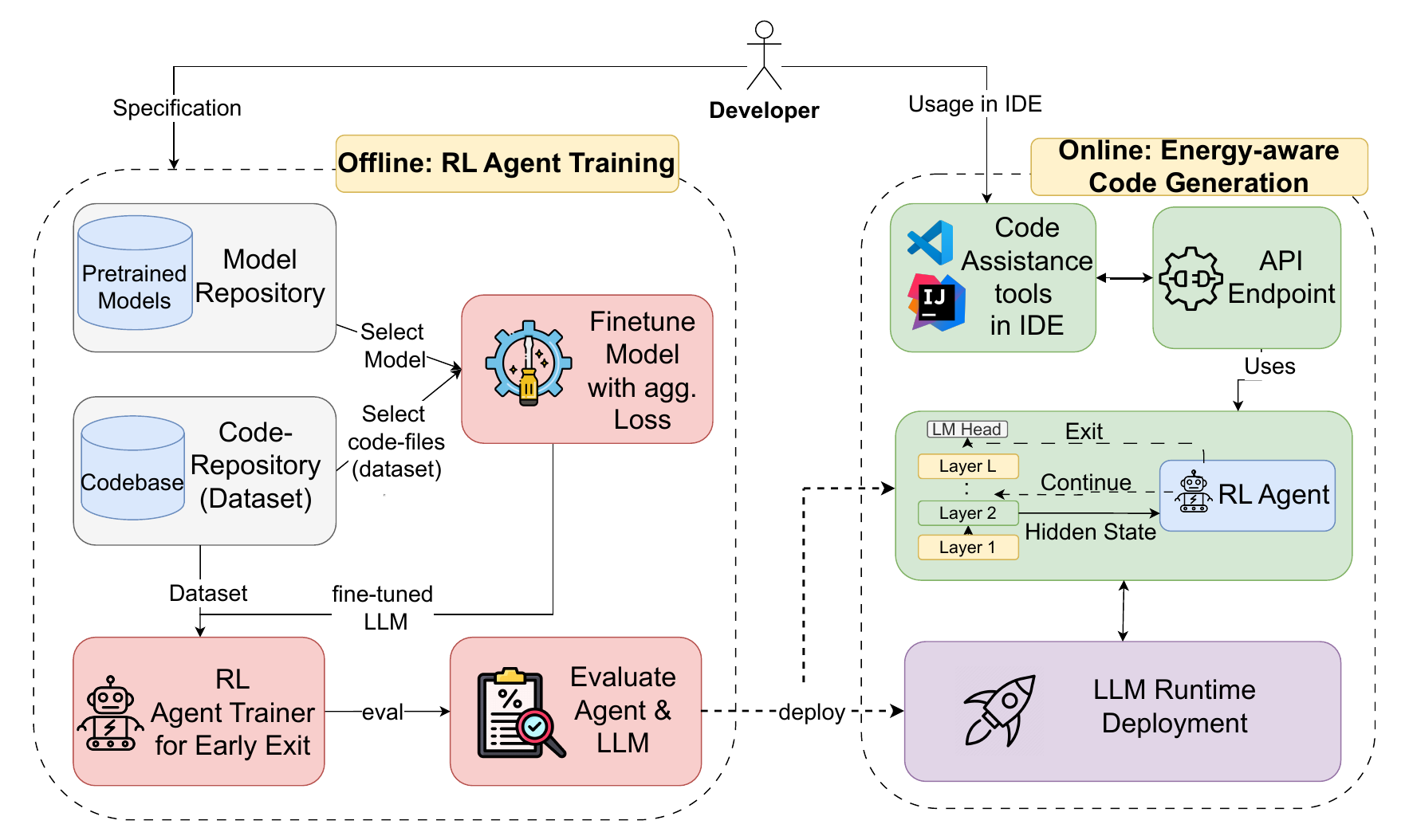}
\caption{A high-level view of the system model.} 
\label{fig:methodology_overview}
\end{figure}

The high-level view of the GREEN-CODE framework is illustrated in our system model as depicted in Figure~\ref{fig:methodology_overview}. It primarily consists of two parts. (1) offline RL agent training, and (2) online energy-aware code generation.

The offline phase of the GREEN-CODE framework involves several steps. \textit{First}, we begin by selecting appropriate pre-trained models and standard input datasets that align with the problem requirements, focusing on real-world code completion tasks.  The selected datasets undergo preprocessing to ensure compatibility with the LLM's input format. This involves tokenization, normalization, and splitting into training, validation, and testing sets, among other processes. \textit{Second}, we apply a specialized fine-tuning process to make LLMs suitable for early exiting. This involves introducing an aggregated, weight-based loss function~\cite{varshney2023}, enabling the model to decode from hidden states of an intermediate layer. \textit{Third}, we formulate the exit method as a Reinforcement Learning (RL) problem. In this setup, the RL agent is trained to learn and balance the trade-off between computational resources used, energy consumption, and output quality (accuracy) by dynamically predicting exit points.  \textit{Finally}, we evaluate RL agent's performance capturing both model-related performance metrics and resource efficiency metrics, like energy consumption.

Once the RL agent's training converges, during the online phase,  the RL agent and the fine-tuned LLMs are deployed on a service end point for runtime usage.
In the following subsection, we discuss each of these components in detail.

\subsection{Data Selection \& Preparation}\label{sec:data}

In this work, we consider on-the-fly code-to-code completion task with whole code files, instead of single function from a given docstring. The goal here is to predict the next token(s) given some code context. Accordingly,  we selected two datasets from the CodeXGlue~\cite{lu2021} benchmark. CodeXGlue provides dataset collection for various code-related tasks. In total, it consists of 10 tasks with 14 datasets. We specifically use the datasets designated for the code completion task, namely, JavaCorpus~\cite{Allamanis2013} and PY150~\cite{Raychev2016}, representing two widely used programming languages, i.e., Java and Python, respectively.

\begin{table}[h] 
\centering
    \resizebox{.95\columnwidth}{!}{%
    \begin{tabular}{llll} 
    \hline
    \textbf{Name} & \textbf{\#Samples/split} & \textbf{\#Tokens} & \textbf{Language} \\
    \hline    \textbf{JavaCorpus~\cite{Allamanis2013}} & 12.934/7.189/8.268 & 15.8m/3.8m/5.3m & Java  \\ 
    \textbf{PY150~\cite{Raychev2016}} & 100.000/50.000 & 76.3m/37.2m & Python  \\
    \hline
    \end{tabular}} 
    \caption{Datasets overview (derived  from CodeXGlue~\cite{lu2021}).}
    \label{method:datasets}
\end{table}

Table~\ref{method:datasets} summarizes the two datasets. We follow the train-test split introduced in CodeXGlue~\cite{lu2021}. Both datasets are collected from open-source projects on GitHub and include thousands of different code files.

\subsection{Candidate Model Selection }
In this work, we considered open-source autoregressive, decoder-only models, as these represent the state-of-the-art LLM models recently released by several large institutions.
We use two model families: OPT~\cite{zhang2022a} (Open Pre-trained Transformers) and Llama~\cite{touvron2023}.
While many variations of these model families exist, distinguished by their size in terms of the number of parameters, we selected smaller-sized open-source models based on our in-lab testbed resources. The main resource constraint is
full-parameter fine-tuning, which requires manifold runtime GPU memory of its model size, for storing optimizers and activations besides the model itself.
With these considerations, we chose OPT and LLaMa 3.2, with parameter sizes of 2.7B and 3B, and consisting of 32 and 28 layers, respectively. Table~\ref{method:models} summarizes the models we chose for the experiments. 
For both models, we use the open-source versions available from Huggingface.

\begin{table}[h] 
\centering
\scriptsize
    \resizebox{0.6\columnwidth}{!}{%
    \begin{tabular}{lll} 
    \hline
    \textbf{Model} &  \textbf{\#Parameters } & \textbf{\#Layers}  \\ 
    \hline
    \textbf{OPT} & 2.7B  & 32 \\ 
    \textbf{Llama 3.2} &3B & 28 \\ 
    \hline
    \end{tabular}} 
    \caption{Models used in this study}
    \label{method:models}
    \vspace{-1\baselineskip}
\end{table}

\subsection{Finetuning Pretrained Models for Early Exiting}\label{sec:finetune}
Standard pretrained LLMs cannot be directly used for dynamic early exits. In typical inference settings, the hidden state from the final layer is passed through a Language Modeling Head (LM head). It is a linear fully connected layer that maps the hidden state to a token prediction within the vocabulary ~\cite{Vaswani2017}. However, in the context of early exiting, when the final layer's hidden state is not utilized, an alternative approach is required to produce the final output. There are two primary strategies for this: (i) training additional LM head at each potential exit layer to map hidden states from that layer, or (ii) employing specific fine-tuning techniques to make intermediate layer decoding viable.

Using multiple LM heads requires significantly higher memory resources. To overcome this,  GREEN-CODE implements specialized fine-tuning to decode from intermediate layers with a single LM head. For this, we adapt the approach presented in~\cite{varshney2023}, which introduces the LITE (Losses from InTermediate layErs) method.
Here, the original LM head is used to collect losses from the intermediate layers. Therefore, the loss function in this fine-tuning process includes intermediate layers:

\begin{equation}
    Loss=\frac{\sum_{i=1}^N w_i\cdot loss_i}{\sum_{i=1}^N w_i}
\end{equation}
where ( $w_i$ ) is a weight given to the loss of layer $i$ ($loss_i$) and $N$ is the number of layers. This loss is then propagated back through the network, enables the model to learn to decode from intermediate layers.

However, fine-tuning a model to have a single LM head for early exit requires other careful considerations as it leads to some potential performance loss. Fine-tuning is a resource- and time-intensive activity, finding optimal weights for different layers (hyperparameters $w_i$ ) is challenging. For instance, giving more weight to lower layers should improve the results in lower layers, however, it leads to an overall performance loss.  Moreover, identifying optimal $w_i$ weights (an NP-hard problem)  and exit points with exhaustive hyperparameter optimization is computationally expensive and even infeasible.

With these considerations, we implemented the following rules to define exit points:
\begin{itemize}
\item The earliest exit point for both models is set at layer 4, ensuring that a portion of the model’s full performance is retained in any case.
\item To facilitate early exits as soon as possible, we allow exits on alternating layers in the first half of the model, skipping every second layer. In the second half of the model, exits are permitted at every fourth layer.
\end{itemize}

This configuration results in 9 exit points for LLaMa (28 layers) and 10 exit points for OPT (32 layers).

Additionally, we assigned the weights $w_i$ as follows: we defined separate budgets ( $\alpha$ ) for the first and second halves of the layers. The budget  $\alpha$  for the first half of layers is set to  $0.7 $, and the second half of layers is allocated $0.2$. The final layer, originally connected to the LM head, received a fixed  $\alpha$  of  $0.1 $ to maintain the ability to make predictions through the last layer. In our settings, we explore only configurations with decaying weights, assigning the highest weight to the earliest exit, with the hypothesis that this would improve performance at shallower layers. This configuration helps to provide more weight to early layers, thereby reducing the inference cost by preferring early exits during runtime.

We use geometric sequences to assign weights, and   ratio for each layer in a group is calculated %
with a decay factor  $r = 0.9$. 
\begin{wrapfigure}{r}{0.5\columnwidth}
\centering
\vspace{-1\baselineskip}
    \includegraphics[width=0.49\columnwidth]{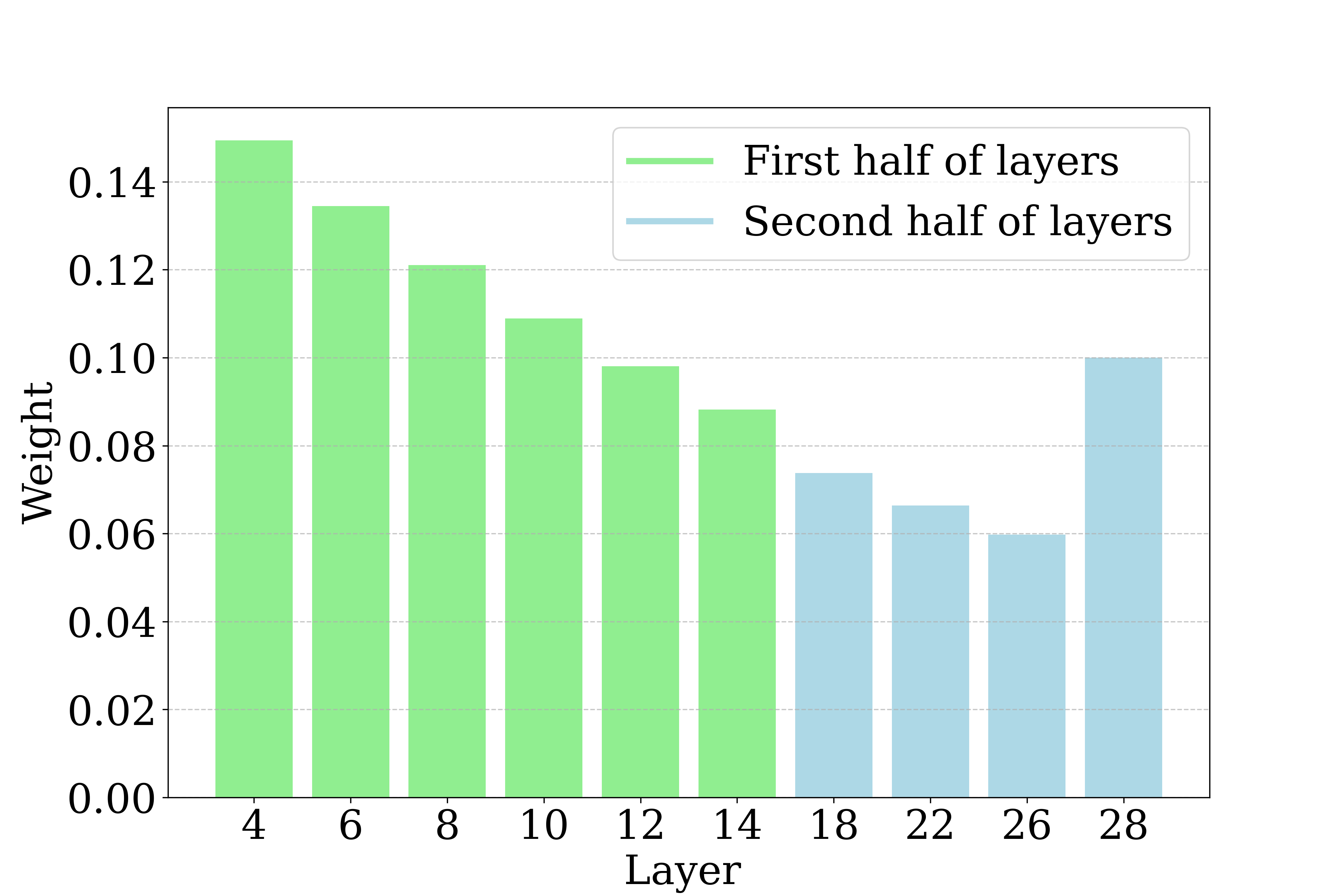}
    \caption{$w_i$ distribution for Llama}
    \label{methodology:weightdist_llama}
\end{wrapfigure}
Next, we normalize each ratio and multiply it by the specified budgets  $\alpha$  (0.7 and 0.2) to compute  $w_i$.

These weight distributions for the Llama 3.2 model are illustrated in Figure~\ref{methodology:weightdist_llama}, where the green bars represent the layers of the first half, and the blue bars represent the layers of the second half.

\textbf{Analysis of fine-tuning method:} 
For both models, we set (the) hyperparameters to the default (values) provided by the Hugging Face training API. Due to the significant difference in the size of the datasets, we adjusted the training process accordingly. Specifically, we trained for only one epoch on the PY150 dataset, which contains approximately 100k samples, compared to JavaCorpus, with around around 15k samples, where we trained for 5 epochs (OPT) and 10 epochs (Llama). Given the large variability in the size of training samples, we split the samples according to a maximum sequence length. We used a learning rate of $1e^{-5}$, a batch size of $4$, and gradient accumulation steps of $32$. Additionally, when necessary, we used packing to collapse small samples together.

The fine-tuning experiments took approximately 24 hours and 19 hours on the Llama and OPT models, respectively (approximately 2-3k steps). We did not perform additional hyperparameter tuning during LLM fine-tuning. However, it is important to note that any additional optimization during fine-tuning, including hyperparameter optimization, would likely enhance the overall performance of the code generation task.

\begin{figure}[h]
\centering
\vspace{-1\baselineskip}
\begin{subfigure}[b]{0.45\columnwidth}
\centering
\includegraphics[width=\textwidth]{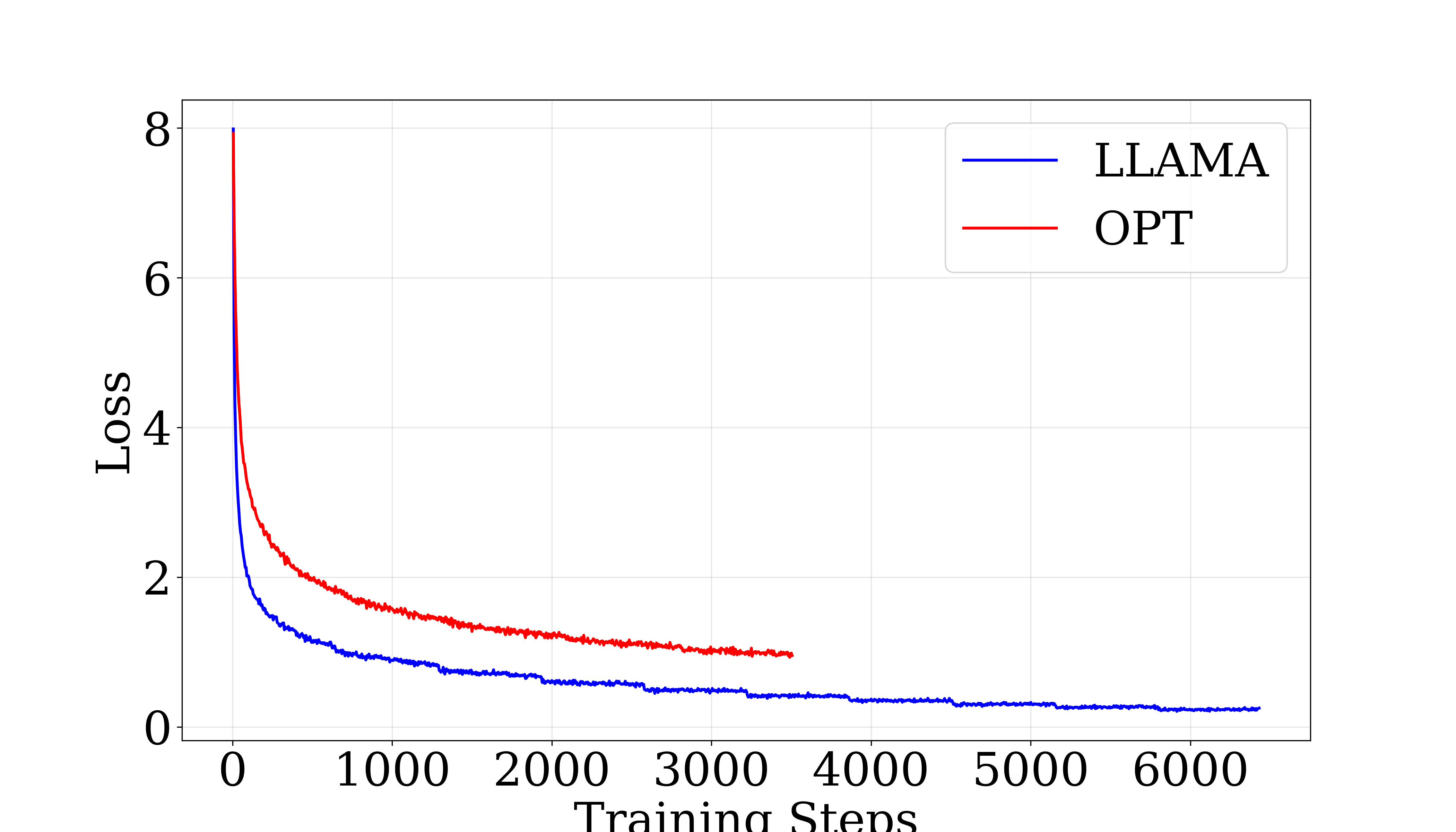}
\subcaption{Loss on JavaCorpus.}
\label{methodology:weight_dist_llama}
\end{subfigure}
\begin{subfigure}[b]{0.45\columnwidth}
\centering
\includegraphics[width=\textwidth]{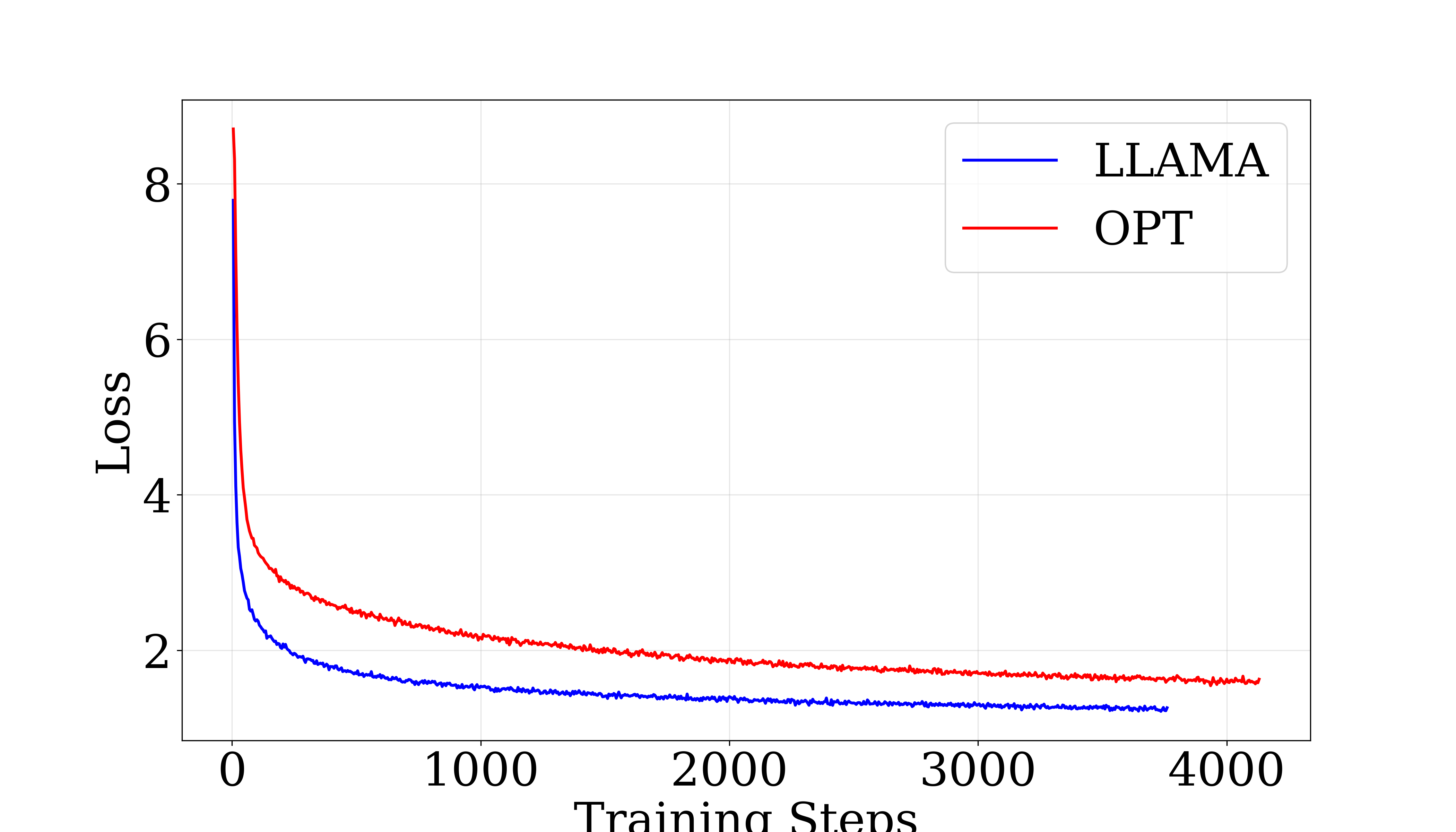}
\subcaption{Loss on PY150.}
\end{subfigure}

\caption{Aggregated Loss of fine-tuned models.}
\label{methodology:finetune_loss}
\vspace{-1\baselineskip}
\end{figure}

Figure~\ref{methodology:finetune_loss} shows the loss curves of both models for two datasets with our fine-tuning method. As observed, after a few thousand training steps, the loss converges and further training has only insignificant effects.
\label{sec:design}
\section{GREEN-CODE: Learning Energy Efficient Code Generation}\label{sec:RL_ee}

The primary objective of our work is to achieve energy-efficient code generation by dynamically selecting exit points during the output token generation. Once we have a fine-tuned model capable of dynamic exiting, we need to develop an algorithm to identify an early exit strategy.

The GREEN-CODE framework formulates the dynamic early exit problem as a reinforcement learning (RL) problem. 
The following subsections describe the individual components of the RL agent's training process.

\subsection{Defining the Environment}
 
\begin{wrapfigure}{r}{0.6\columnwidth}
\vspace{-1\baselineskip}
\centering
    \includegraphics[width=0.6\columnwidth]{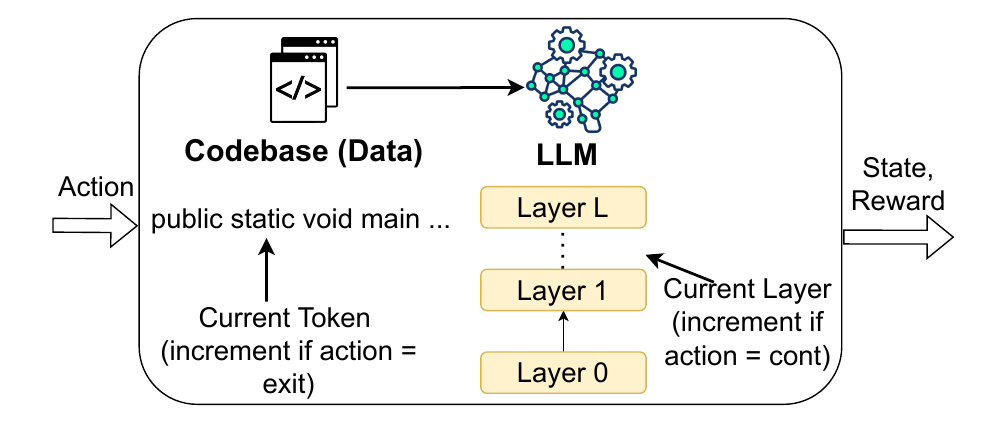}
 \caption{Illustration of the RL environment.} 
\label{fig:environment_overview}
\vspace{-1\baselineskip}
\end{wrapfigure}
One of the main components of formulating any problem into the RL domain is to create an environment where an agent can interact and learn the complexities and trade-offs between different configurable and dependent parameters and take decisions accordingly.
A high-level overview of the environment is illustrated in Figure~\ref{fig:environment_overview}. It comprises two main components. (i) \textit{Codebase,} a dataset that the LLM processes to generate observations.
(ii) \textit{Fine-tuned LLM model}, capable of exiting at intermediate layers (see Section \ref{sec:finetune}).

\subsection{Defining State Space}
In RL, a state represents the current configuration of the environment. This state, comprising multiple features, provides all the necessary information for the agent to take action to achieve its goals. However, including excessive features increases the state space exponentially~\cite{Sutton1998}, making the training and use of RL agents computationally expensive. Particularly in our code generation case, the inference must be performed in real time with minimal computational overhead.

Therefore, we use the \textbf{\textit{current layer hidden state}} produced by the last layer. This hidden state encapsulates the current contextual understanding of the layer, and it does not require additional computational and monitoring overhead to gather from the environment.

\subsection{Designing an Action Space}
An action is a decision taken by an RL agent to transition between states in its environment. The agent selects actions based on a policy, which is a strategy mapping states to actions, to maximize cumulative rewards over time. For our dynamic early exit problem, we define two concrete actions:\textbf{ \textit{ (1) 
 continue,} and (2) \textit{exit}.}

When the RL agent decides to \textit{exit} at the current layer $l$, the token currently being processed is considered complete. 
The agent receives the environment state as the hidden state of the next token at the first exit point, i.e., the first hidden state of the next token, along with the reward.

When the RL agent decides to \textit{continue},  the environment remains at the current token but advances to the next layer. The agent then receives the environment state as the hidden state of the new current layer. When the environment reaches the last layer, we mark the current token as finished and proceed with the exit action. %
Actual exit points, i.e., the layers at which to exit, are determined using exit points from the fine-tuning method discussed in Section~\ref{sec:finetune}.

\subsection{Reward Function}
Rewards give feedback on the value of an agent's actions. Upon taking an action, the agent receives a reward signal from the environment, indicating how beneficial the action was for achieving the desired outcome. Thus, accurately defining rewards is crucial for shaping the agent's learning effectively.

The intuition of this paper and empirical observations (Section~\ref{sec:motivation}) indicate that many tokens can be accurately predicted with just a few early layers. Thus, an agent should exploit this property efficiently by taking advantage of early exits to reduce energy consumption. However, in many cases, deeper layers are necessary to avoid degraded accuracy. Consequently, rewarding an RL agent for aggressive early exits leads the agent to avoid deeper layers completely. Hence, we need a reward structure to balance this trade-off.
Additionally, we need to consider LLM-specific intricacies while designing reward structures. LLMs process input one token at a time, so using episodic rewards (i.e., rewards based on performance over an entire sample) can mislead the agent. This is because a single token's poor reward can overshadow the correct actions taken for other tokens. Therefore, the agent should make decisions based on local properties while optimizing for global, sample-level consistencies, as rewards are optimized over an episode.

Based on these insights and observations, we defined our reward functions for energy efficient early-exiting as below:
\begin{equation}
r_e =
\begin{cases}
    \text{1}, & \text{if } y_{\text{pred}} = y \text{ and } \ell_{\text{curr}} = \ell_{\text{opt}} \\
    -(\ell_{\text{curr}} - \ell_{\text{opt}}) \cdot \alpha, & \text{if } y_{\text{pred}} = y \text{ and } \ell_{\text{curr}} \neq \ell_{\text{opt}}\\
    
    -(\ell_{\text{opt}} - \ell_{\text{curr}})  \cdot \beta, & \text{if } y_{\text{pred}} \neq y \text{ and } \ell_{\text{curr}} < \ell_{\text{opt}} \\
    
    -\epsilon, & \text{otherwise } 
\end{cases}
\label{eq:reward1}
\end{equation}
\begin{equation}
r_c =
\begin{cases}
    \text{1}, & \text{if } \ell_{\text{curr}} < \ell_{\text{opt}} \\
 
    -(\ell_{\text{next}} - \ell_{\text{opt}}) \cdot \gamma, & \text{otherwise}
\end{cases}
\label{eq:reward2}
\end{equation}

Here, $\ell_{curr}$ is the current layer being considered, $\ell_{opt}$ is the first layer whose prediction matches the prediction of the final layer (the optimal exit point), and $\ell_{next}$ represents the subsequent layer. $y$ denotes the token prediction at the last layer, serving as the ground-truth token for reward calculation, while $y_{pred}$ represents the prediction at $\ell_{curr}$. The coefficients $\alpha, \beta, \gamma$ are trade-off parameters, constrained by $0 \leq \alpha, \beta, \gamma \leq 1$, which controls the agent's learning behavior. We also scale penalties to the interval $[-1, 0]$ to stabilize learning. For simplicity, we represent all layers in the formulation; however, our specific exit points are based on fine-tuning method (i.e., exits are not allowed on non-finetuned layers), and rewards are calculated accordingly. 

The possible reward for an \textbf{exit action $r_e$} is given in Equation~\ref{eq:reward1}. Here, \textit{\textbf{the first condition}} applies a fixed reward if the decision to \textbf{exit} is optimal, meaning the current layer's prediction is correct (i.e., matches the prediction of the final layer), and it is the shallowest layer to do so, thus, $\nexists \ell' | y' = y \land \ell' < \ell_{curr}$.
\textit{\textbf{The second condition}} in Equation~\ref{eq:reward1} is the penalty for exiting too late. It occurs when the current layer produces a correct token, but a shallower layer $\ell'$ exists such that $\ell' < \ell_{curr}$ with $y' = y$. This penalty is proportional to the distance between $\ell_{curr}$ and $\ell_{opt}$, scaling with the number of steps from the optimal layer. While such late exits still yield correct predictions in practice, overly rewarding them causes the agent to exploit exiting decisions excessively. Thus, the agent should avoid making  suboptimal exits. Nevertheless, in experiments, we set $\alpha \le \beta$ so that exiting late is at least as good (or better) than exiting too early. \textit{\textbf{The third condition}} in Equation~\ref{eq:reward1} represents the worst-case scenario: exiting at a layer that is too early, resulting in an incorrect prediction. The penalty here is also scaled based on the distance from the optimal exit layer.
The \textit{\textbf{final condition}} for exit,  covers rarely occurring edge cases where $y_{pred} \ne y$ and $\ell_{curr} > \ell_{opt}$. We assign a small constant penalty to this scenario.

Similarly, the possible reward for \textbf{continue action $r_c$} is given in Equation~\ref{eq:reward2}. Here, we reward optimal decisions with a constant value of $1$ when continuing. A continuation is considered optimal if $\ell_{curr} < \ell_{opt}$, meaning the agent has not yet reached the optimal layer. In contrast, we penalize all other cases where $\ell_{curr} \geq \ell_{opt}$, scaling the penalty based on the distance between the current layer and the optimal layer. Here, we use $\ell_{next} = \ell_{curr} + 1$ because a continuation is incorrect if $\ell_{curr} = \ell_{opt}$, since the agent should have exited at this point. Again, we introduce a coefficient to control the penalty. Increasing coefficients $\alpha$ and $\gamma$ should lead the agent to avoid too late exits, while increasing $\beta$ avoids too early exits, managing the trade-offs.

\subsection{Cost Function}
The objective of the training is to maximize the objective function in Equation~\ref{eq:objective}. Here,  $r_c$ and $r_e$ are step rewards as defined above. Note that we define $r_{c}=0$ if the action is exit and $r_{e}=0$ if the action is continue.

\begin{equation}\label{eq:objective}
J(\theta) = \mathbb{E}_{\pi_{\theta}} \left[ \sum_{t=1}^{T} \left(  r_e(t) +  r_c(t) \right) \right]
\end{equation}

where $\pi_{\theta}$ is the  policy  parameterized by $\theta$, $T$ is the number of time steps per episode and $t$ is the current time step.

\subsection{Training an RL agent}
A RL agent needs to be trained to understand the complexities of the environment and take appropriate actions to achieve the desired objective. In this case,  managing trade-offs between energy consumption, the accuracy of the code generation task and identifying optimal exit points. For this, RL agent requires multiple episodes during the training period and attempts to maximize its rewards for actions taken in each episode. Once the cumulative reward across multiple episodes stabilizes, we assume the RL agent has converged.

In each episode, the RL agent performs resets and steps within the environment. A reset can occur either when an exit action is taken on the last token predicted by the  LLM or when the last layer of the final token is surpassed by a continue action. During a reset, the environment first samples a code file uniformly from the dataset and generates $T$ tokens using the LLM. The context provided to the LLM is determined randomly by uniformly sampling from the interval $[0.2, 0.6]$.
If an exit action is taken, the internal state of the environment moves to the next token, and rewards are assigned accordingly. If a continue action is taken, the process advances to the next layer, mirroring the way early exiting is used during inference.

In summary, we provide an environment that gives the agent only the hidden state of the current layer and current token. 
Therefore, agents must learn the  efficiency-accuracy trade-off solely based on the hidden state and the reward signal.

\label{sec:methodology}
\section{Implementation and Deployment}

We employed the standard Gymnasium~\cite{towers2024} framework to implement our GREEN-CODE solution. It is developed based on the OpenAI Gym~\cite{brockman2016} framework. Gymnasium provides an API that facilitates communication between learning agents.
For RL algorithms, we used the widely adopted PyTorch-based  Stable-Baselines3 (SB3) library~\cite{Raffin2021}, which  provides the implementation of state-of-the-art RL algorithms.

As a specific algorithm, we chose the Proximal Policy Optimization (PPO)~\cite{schulman2017} algorithm, an RL method that has demonstrated effectiveness across various tasks in different domains. PPO offers stability in training by using clipping approaches to avoid excessively large updates, leading to more stable training. We employed single environment and shallow network configurations. We trained the network until convergence was observed in both episodic and stepwise rewards. On average, it took 200k-500k steps before observing convergence. A detailed description of the hyperparameter setup and results are provided in the next section.

\textbf{Inference Deployment:} We extract the policy from the trained RL agent and convert it into a PyTorch-based Deep Neural Network (DNN) by utilizing the corresponding policy and action networks from SB3.

For deployment, we implemented an endpoint that deploys the LLM alongside an RL agent. Consequently, our setup can be utilized similarly to GitHub Copilot or other tools. Furthermore,  
we also integrated GREEN-CODE with an existing VSCode extension from Huggingface. The endpoint adheres to the Huggingface Inference API, enabling our LLMs to be used in the Huggingface VSCode extension\footnote{https://github.com/huggingface/llm-vscode}. Using our deployed LLM endpoint and extended configurations of the  VSCode extension, we are able to specify runtime LLM behavior (e.g., \texttt{num\_predict}) as well as RL agent thresholds to decide trade-off between resource consumption and accuracy. 
We implemented our framework using Python 3.12 programming language.
GREEN-CODE prototype implementation,  deployment configuration scripts, and details, including libraries and dependencies used, are available in an open-source GitHub repository at \textbf{\href{https://github.com/Large-scale-Sustainable-Computing-LSC/green-code}{https://github.com/Large-scale-Sustainable-Computing-LSC/green-code}}.

\label{sec:implementation}

\section{Performance Evaluation}

\subsection{Experimental Setup}
We used an in-lab GPU server. 
Our machine is equipped with an NVIDIA Quadro RTX 8000  as the GPU accelerator, with  49152 MiB of VRAM. The server itself is equipped with an AMD EPYC 7452 CPU with 32 CPU cores, 1.0 TiB of main memory, running Ubuntu 20.04.6 LTS. 

We measure the  metrics capturing both efficiency (hardware and model level) and performance of the model itself. 

\subsubsection{Efficiency Metrics}
 We measure the following efficiency-related metrics. 
\textbf{Number of layers skipped,} which is hardware-independent and measures how many layers are skipped during inference. \textbf{Latency,} measures the time (in seconds) taken to get a response for a user request. It is dependent on the hardware used. \textbf{Energy Consumption (in Ws),} provides the real-world cost and sustainability impact of running a model. Unlike theoretical measures, energy consumption accounts for the actual operational power usage cost during inference, reflecting the efficiency of hardware utilization and overheads introduced. Finally, \textbf{Throughput,} measures the number of tokens/second generated, providing a fine-grained analysis of model performance.

We directly measure number of layers skipped, throughput, and latency. Energy consumption is measured using ZeusMonitor \footnote{https://ml.energy/zeus/}, which uses \texttt{pynvml}, but also approximates for a lower sampling rate.

\subsubsection{Performance  metrics}
Traditional NLP evaluation often focuses on text matching through metrics like $n$-gram matching, code generation demands additional syntactic and semantic correctness to ensure functionality. 
Standard LLM metrics like\textbf{ BLEU} and \textbf{ROUGE},  provide a broad performance estimate~\cite{Evtikhiev2023},  prioritizing lexical precision, which may not align with the evaluation of the functional aspects of the code. Accordingly, we also use code generation specific metric, i.e.,  CodeBLEU~\cite{ren2020}. \textbf{CodeBLEU} still uses $n$-gram matches (similar to BLEU) but also incorporates elements relevant for evaluating code models. We also report ``Syntax" and ``dataflow" metrics, which are sub-metrics of CodeBLEU, enhancing our evaluation. For all these three metrics, the minimum value is 0 and the maximum value is 1.

\subsection{Hyperparameter Settings}

For our RL agent training, we used the hyperparameters depicted in Table~\ref{exper:hyperparams_PPO}. We relied on standard parameters provided by the RL library for training PPO. We limited the network hidden size to 32 or 64 and used 1 or 2 layers to minimize inference overhead, ensure faster training, and reduce the risk of overfitting due to an excessively large network. It is important to note that we did not perform hyperparameter optimization due to the significant computational costs associated with it. Nevertheless, our empirical results still demonstrate superiority over baselines, and further hyperparameter optimization would likely enhance performance.

In our evaluation, we use the softmax output of the policy network to decide which action to take. By adjusting the temperature parameter ($T$) and softmax thresholds, we can control the balance between exploration and exploitation during inference.
In other words, these two thresholds determine how "sure" the agent must be before deciding to exit (e.g., a threshold of 0.9 is much stricter than a threshold of 0.5). Thus, we report results with different thresholds.

\begin{table}[!t] 
\centering
    \resizebox{\columnwidth}{!}{%
    \begin{tabular}{lll} 
    \toprule
    \textbf{Hyperparameter} & \textbf{Value}  & \textbf{Description} \\ 
    \midrule
    \texttt{steps} & 500000 & Number of training steps. \\
    \texttt{batch\_size} & 512,32 & Number of experiences per gradient update. \\
    \texttt{Buffer Size} & 4096, 256 & Number of experiences collected before gradient update \\
    \texttt{Epochs} & 6,2 & Number of epochs through the buffer when updating gradients. \\
    
    \texttt{learning\_rate} & 5e-5, 1e-4 & Learning rate. \\

    \texttt{lr\_scheduler\_type} & \texttt{linear} & Linear learning rate scheduled over the steps. \\
    \texttt{$\gamma$} & 0.99 & Discount factor \\
    \texttt{Number of hidden layers} & 2 &  How many hidden layers to use\\
    \texttt{Number of hidden units} & 32, 64 & How many hidden units in each hidden layer to use \\
    \bottomrule
    \end{tabular}} 
    \caption{Hyperparameters for PPO training.}
    \label{exper:hyperparams_PPO}
    \vspace{-1\baselineskip}
\end{table}

\subsection{Experiment Design}

In our experiments, we primarily focused on line-completion tasks. Given a code context, the objective is to complete the next line of code (LOC). We set the maximum number of tokens generated (\textit{max\_new}) to 15, ensuring that a full LOC is completed. The average output length for line-level completion is approximately 7 tokens according to CodeXGlue, and some earlier work also uses 10 tokens~\cite{Sun2024}. We split the code file uniformly across samples, using the first 20\% of the tokens of a sample as context. This results in variable context lengths for different samples due to variations in dataset files. However, we limited the maximum context length to 512 tokens to speed up evaluations and avoid memory constraints. Similar thresholds are used CodeXGlue benchmarks are 488 for PY150 and 365 for JavaCorpus.

We provide the model with the first $n$ tokens of a sample as input and use the tokens from the dataset in the range $[n+1, n+max_{new}]$ as ground truth labels. We then calculate corpus-level metrics by comparing the ground truth tokens against the tokens generated by the model. We always evaluate on 1000 samples from the test sets of the datasets.

\subsection{Evaluation of RL Agent Training}\label{perf:rl_agent_eval}
\begin{figure}[h]
\centering

\begin{subfigure}[b]{0.45\columnwidth}
\centering
\includegraphics[width=\textwidth]{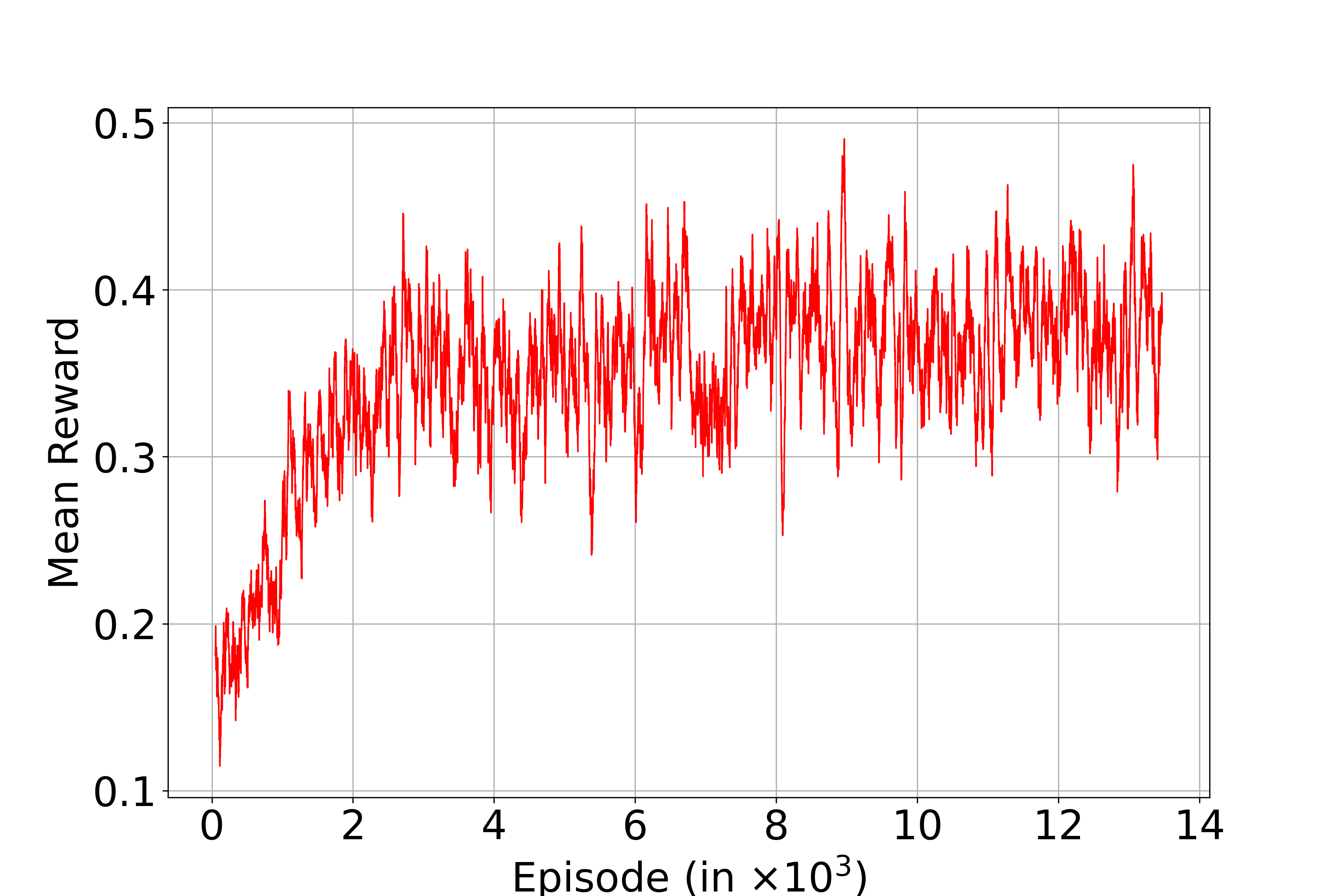}
\subcaption{OPT, JavaCorpus }
\end{subfigure}
\begin{subfigure}[b]{0.45\columnwidth}
\centering
\includegraphics[width=\textwidth]{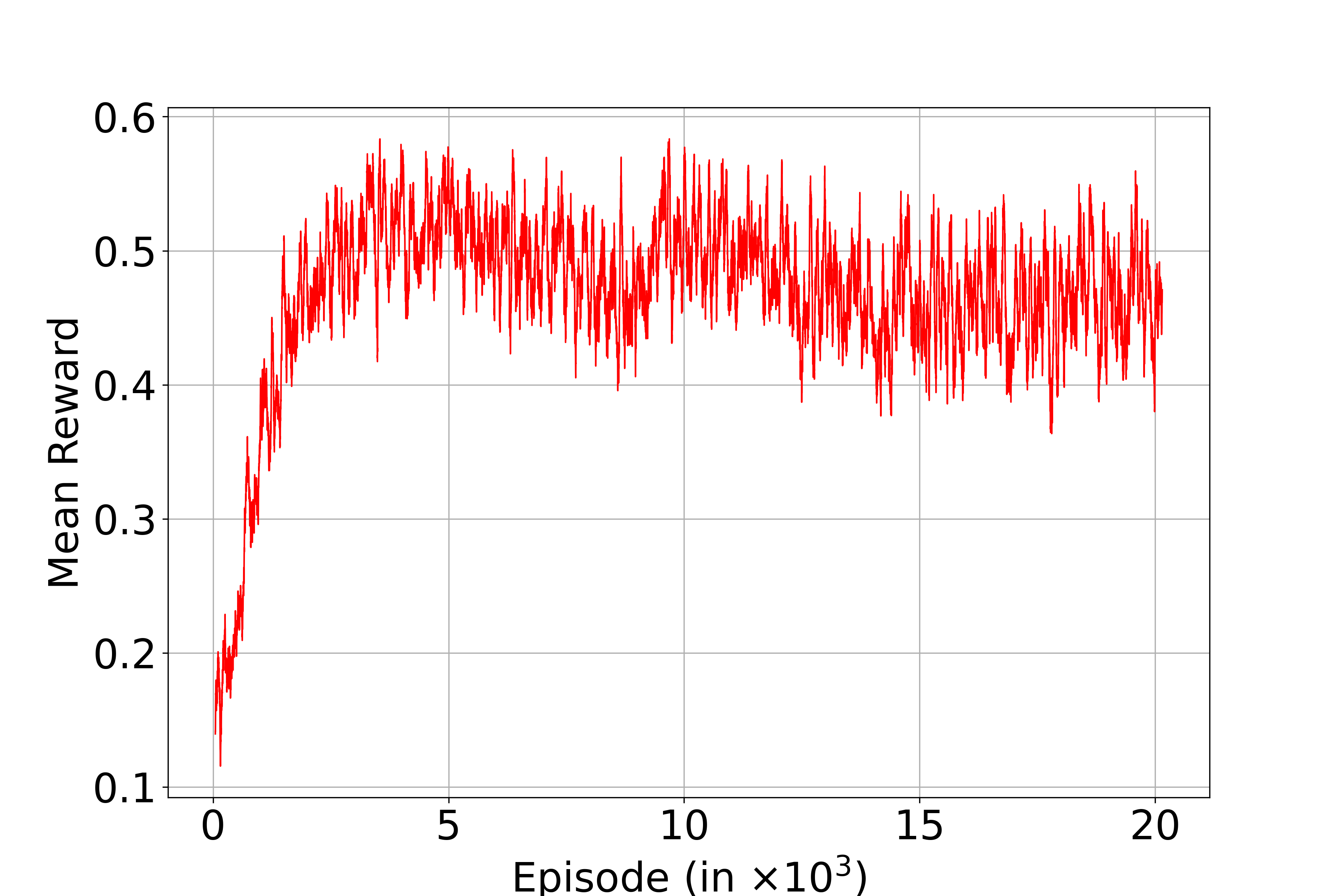}
\subcaption{Llama, JavaCorpus }
\end{subfigure}

\vspace{1em} 

\begin{subfigure}[b]{0.45\columnwidth}
\centering
\includegraphics[width=\textwidth]{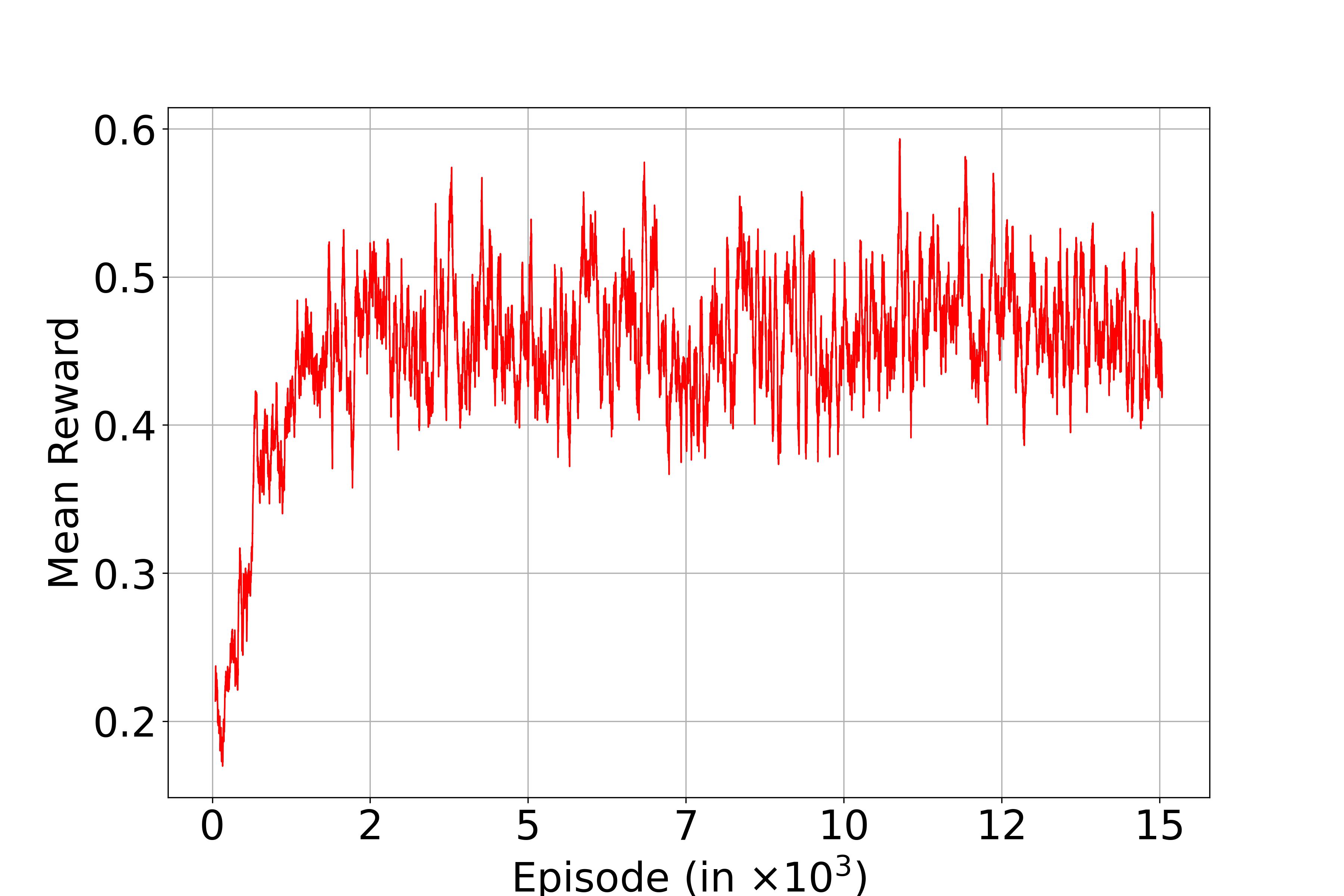}
\subcaption{OPT, PY150 }
\end{subfigure}
\begin{subfigure}[b]{0.45\columnwidth}
\centering
\includegraphics[width=\textwidth]{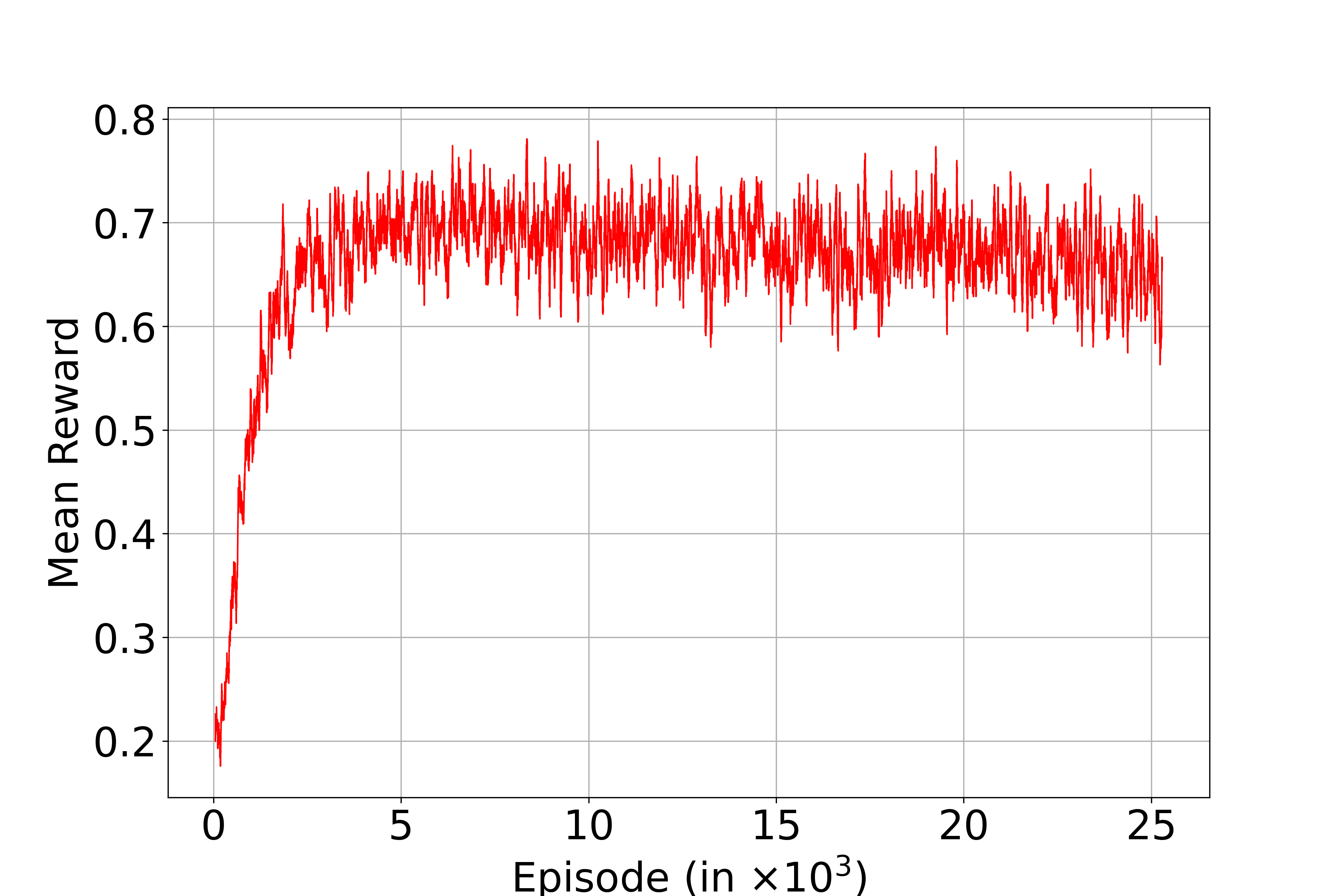}
\subcaption{Llama, PY150}
\end{subfigure}

\caption{Mean step reward per episode of PPO training with Llama and OPT (moving average over 50 episodes).}
\label{experiment:rew_rl_agent}
\end{figure}

Figure~\ref{experiment:rew_rl_agent} illustrates the mean reward per step in each episode for PPO agents trained over 500,000 steps. Note that the number of episodes (x-axis) varies across models due to differences in episode lengths, which depend on the specific model and dataset (e.g., OPT has more "exit points" than Llama, leading to different action choices). For the JavaCorpus dataset, we set $\beta$ and $\gamma$ to 1. For the PY150 dataset, we set $\beta = \gamma = 0.5$, as specified in Section~\ref{sec:RL_ee}, reflecting the higher token-level accuracy of lower layers achieved on this dataset.

As seen in Figure~\ref{experiment:rew_rl_agent}, after the initial few thousand iterations, agents receive exponential cumulative rewards, which stabilize around 2000–2500 episodes across all settings, indicating the convergence of the RL agent.

\begin{figure}[h]
\centering

\begin{subfigure}[b]{0.49\columnwidth}
\centering
\includegraphics[width=\textwidth]{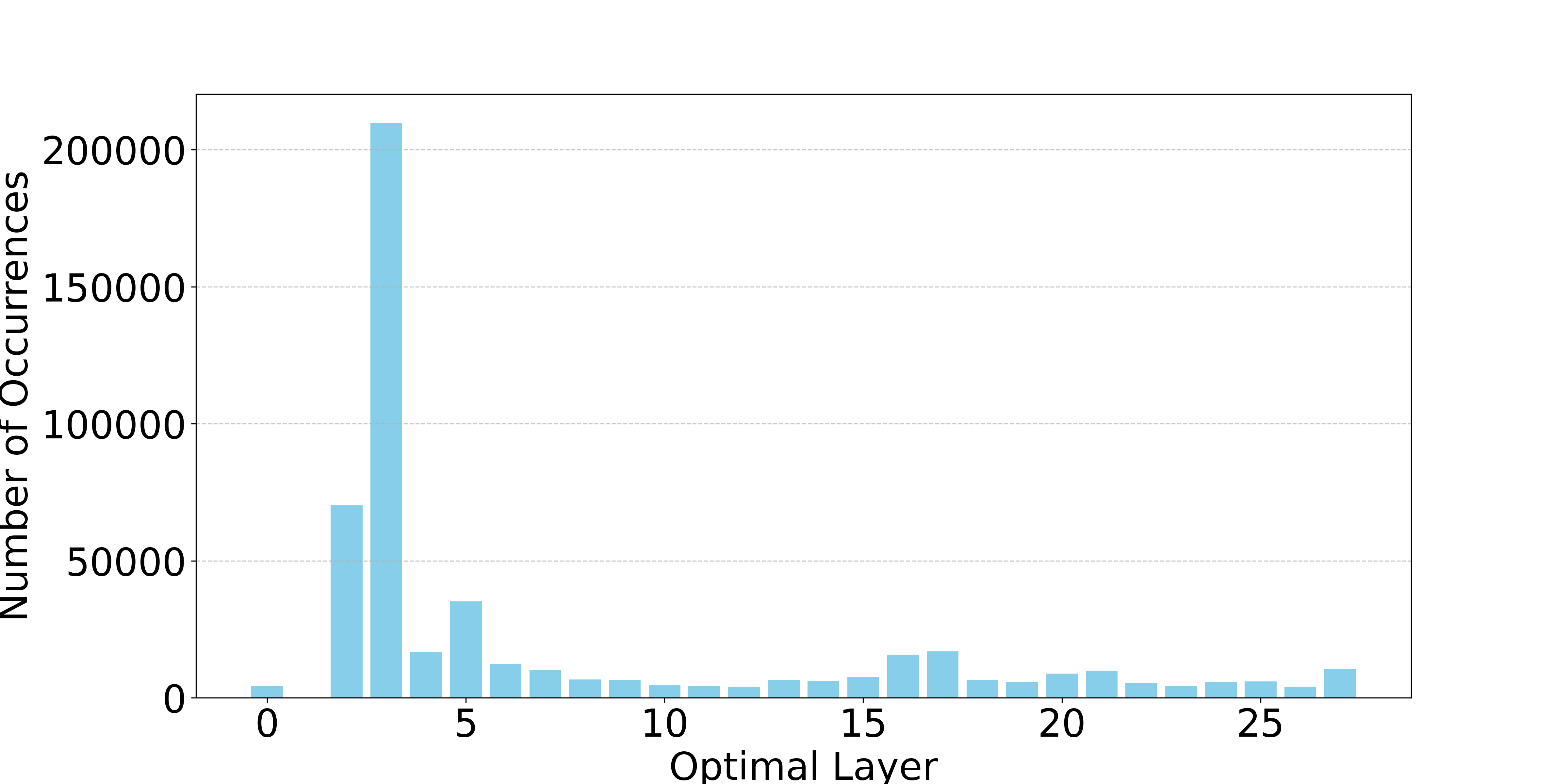}
\subcaption{LLama.}
\label{rl_train_occ_llama_java}
\end{subfigure}
\begin{subfigure}[b]{0.49\columnwidth}
\centering
\includegraphics[width=\textwidth]{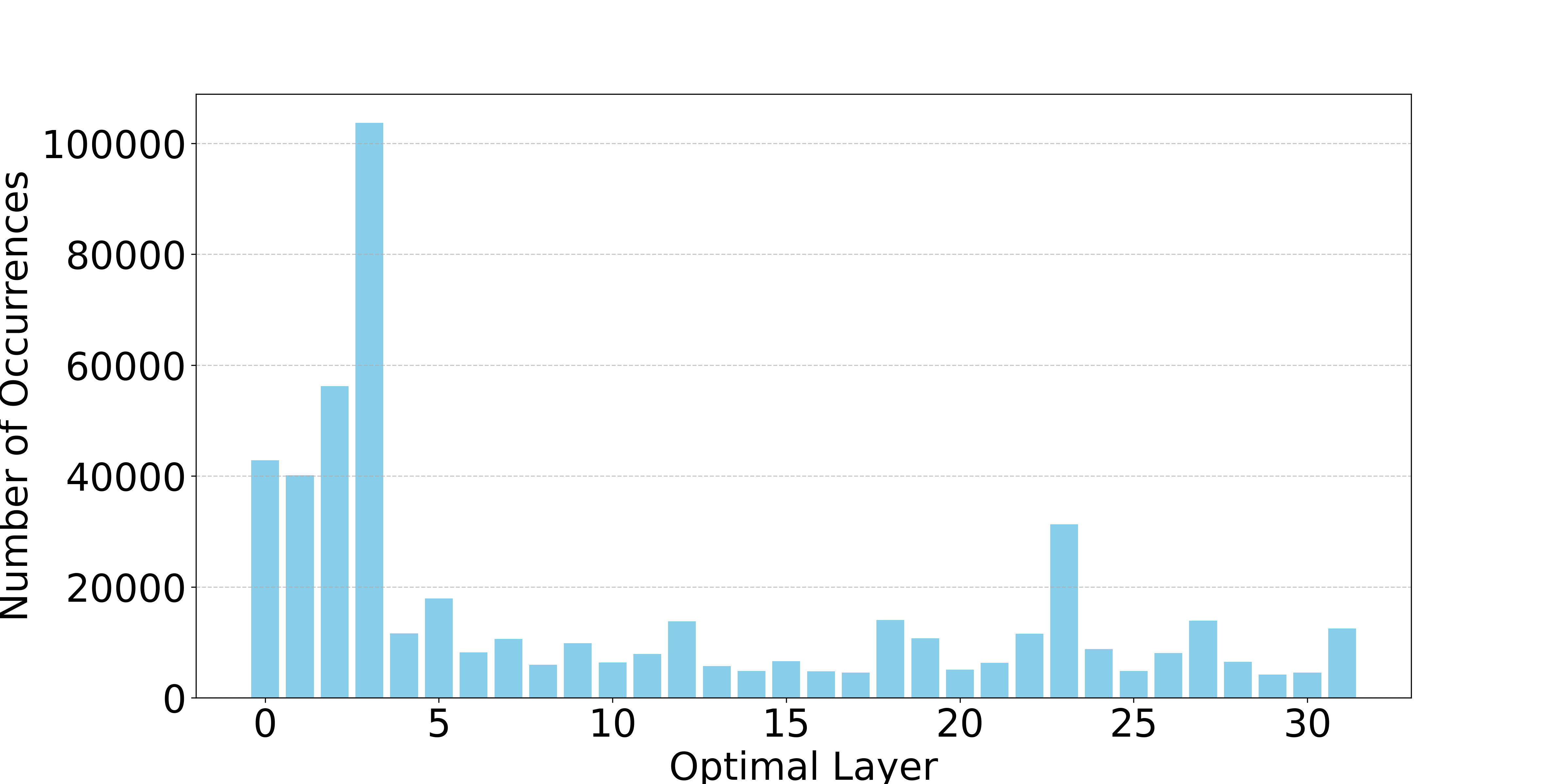}
\subcaption{OPT.}
\end{subfigure}

\caption{Occurrences of optimal exits during training of RL agent (JavaCorpus).}
\label{experiment:rl_train_occurrences}
\vspace{-1\baselineskip}
\end{figure}

Figure~\ref{experiment:rl_train_occurrences} illustrates the optimal exits for all samples encountered during the RL agent training. An optimal exit is defined as the first layer that aligns with the last output layer.

The figure reveals that, with Llama, earlier layers (i.e., 59\%) for optimal exits within the first 5 layers) are more frequently optimal compared to OPT (50\%). This difference may influence learning dynamics. A potential reason for this is the disparity in the number of layers: OPT has 32 layers, while Llama has only 28, despite having a higher total parameter count. As a result, during fine-tuning, OPT may distribute weight across more layers, leading to less emphasis on earlier ones. Consequently, there is a shift when training the RL agents, and the Llama model is more capable overall.

\subsection{Evaluation of Code Generation}
In this section, we present runtime evaluations of our method, which integrates the RL agent into the inference pipeline to dynamically decide when to do early exiting. We use the fine-tuned LLMs capable of early exiting for this. We begin by analyzing the performance on both datasets, and discuss overhead of GREEN-CODE.  

We compare our GREEN-CODE  with  \textbf{two baselines}, (i) \textit{base model}-   the non-fine-tuned version and the (ii) \textit{fine-tuned model}-  which uses all layers, i.e., without early exits. Our proposed GREEN-CODE methods solution is represented as ($GC(T)$), where $T$ is  the temperature threshold.

\subsubsection{Performance Evaluation on JavaCorpus Dataset}
\begin{figure}[h]
\centering

\begin{subfigure}[b]{0.9\columnwidth}
\centering
\includegraphics[width=\textwidth]{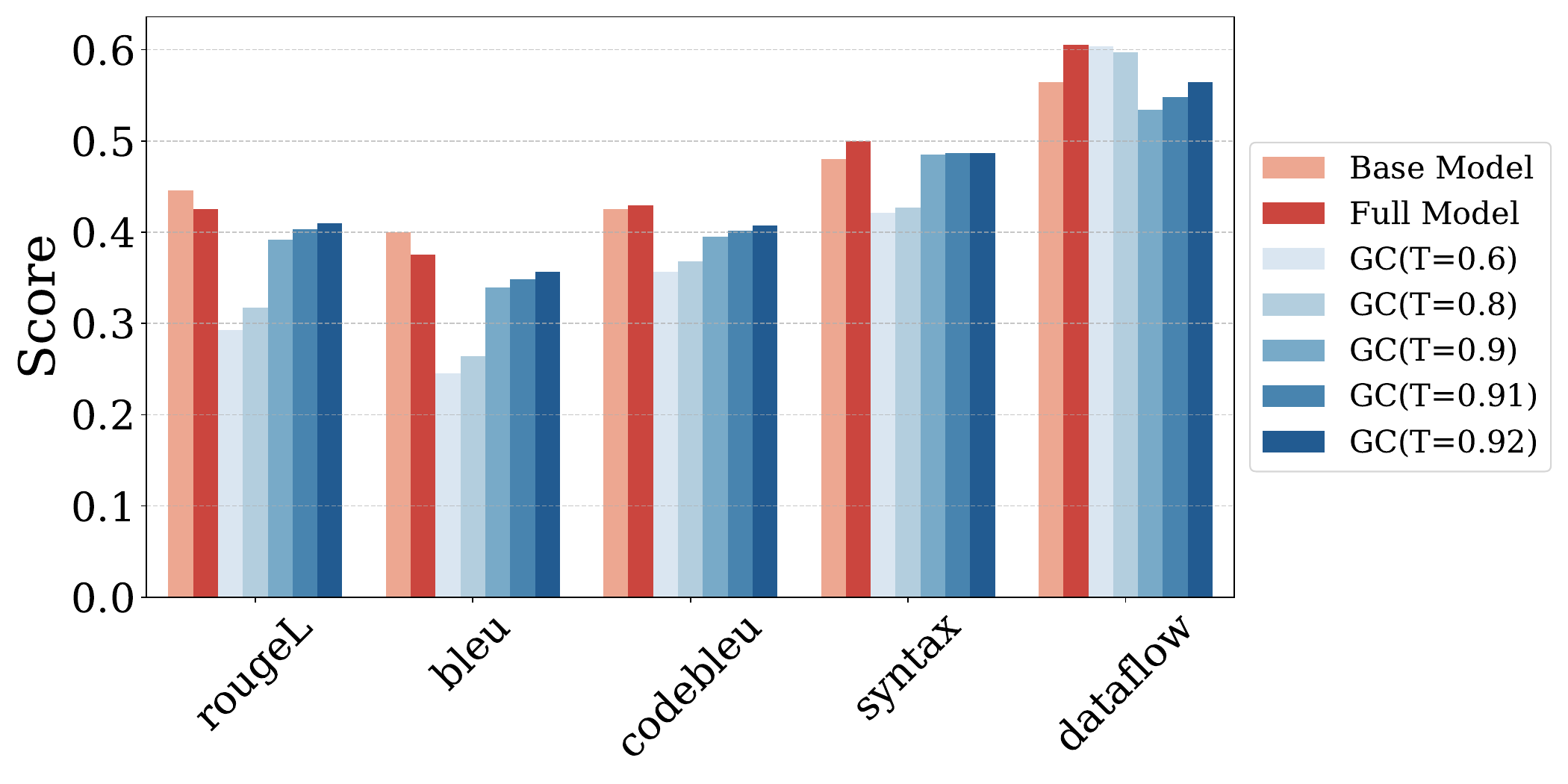}
\subcaption{Score metrics (RougeL, CodeBleu, Bleu)}
\label{llama_java_corpus_metrics:metrics}
\end{subfigure}
\begin{subfigure}[b]{0.95\columnwidth}
\centering
\includegraphics[width=\textwidth]{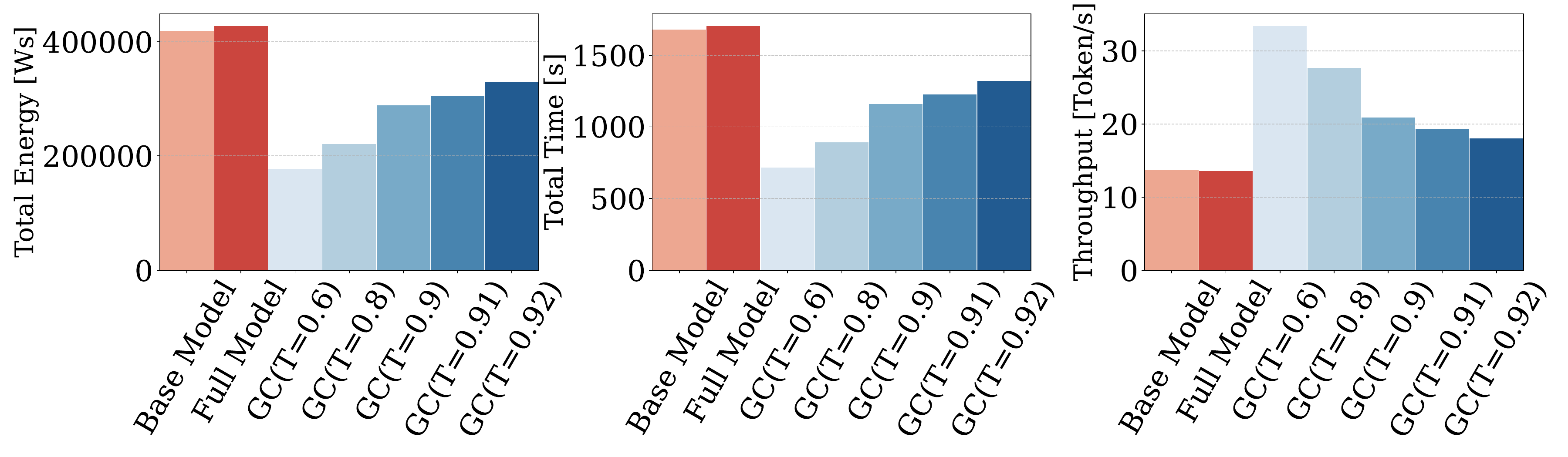}
\subcaption{Resource Consumption metrics (Energy, Time, Throughput)}
\label{llama_java_corpus_metrics:resources}

\end{subfigure}

\caption{Results of Llama on JavaCorpus, with different GC agent thresholds $T$ and baselines. }
\label{experiment:llama_java_corpus_metrics}
\end{figure}

Figure~\ref{experiment:llama_java_corpus_metrics} presents model performance (Figure~\ref{llama_java_corpus_metrics:metrics}) and efficiency metrics, i.e., resource consumption (Figure~\ref{llama_java_corpus_metrics:resources}). 

We observe that the base model performs slightly better than the fine-tuned one in terms of accuracy for all metrics except CodeBLEU, where the fine-tuned model performs slightly better due to improved syntax and dataflow scores. Furthermore, the energy/time difference between these two models is negligible, as expected, since they use the same number of layers.

However, our approach, Green-Code (GC)-  dynamic early exit with RL agent performs well with low exit thresholds. For example, with $T=0.6$, the GC achieves a RougeL score of about 0.29, while the full model achieves about 0.42, using less than half the energy or time. Additionally, accuracies can be increased by trading off some computational resources. For instance, in the most aggressive setting ($T=0.92$), the GC-model achieves a RougeL score of about 0.41, while the full model achieves around 0.425, while saving ~23\% in energy and time. Moreover, the throughput of the base models is roughly 13 tokens/second, compared to 18 tokens/second in our GC($T=0.92$). With a more balanced threshold of $T=0.9$, we can still achieve good performance (i.e., RougeL of 0.39) while saving significant energy and latency costs. Therefore, our approach enables dynamic requirements based on developer needs.

\begin{figure}[h]
\centering

\begin{subfigure}[b]{0.9\columnwidth}
\centering
\includegraphics[width=\textwidth]{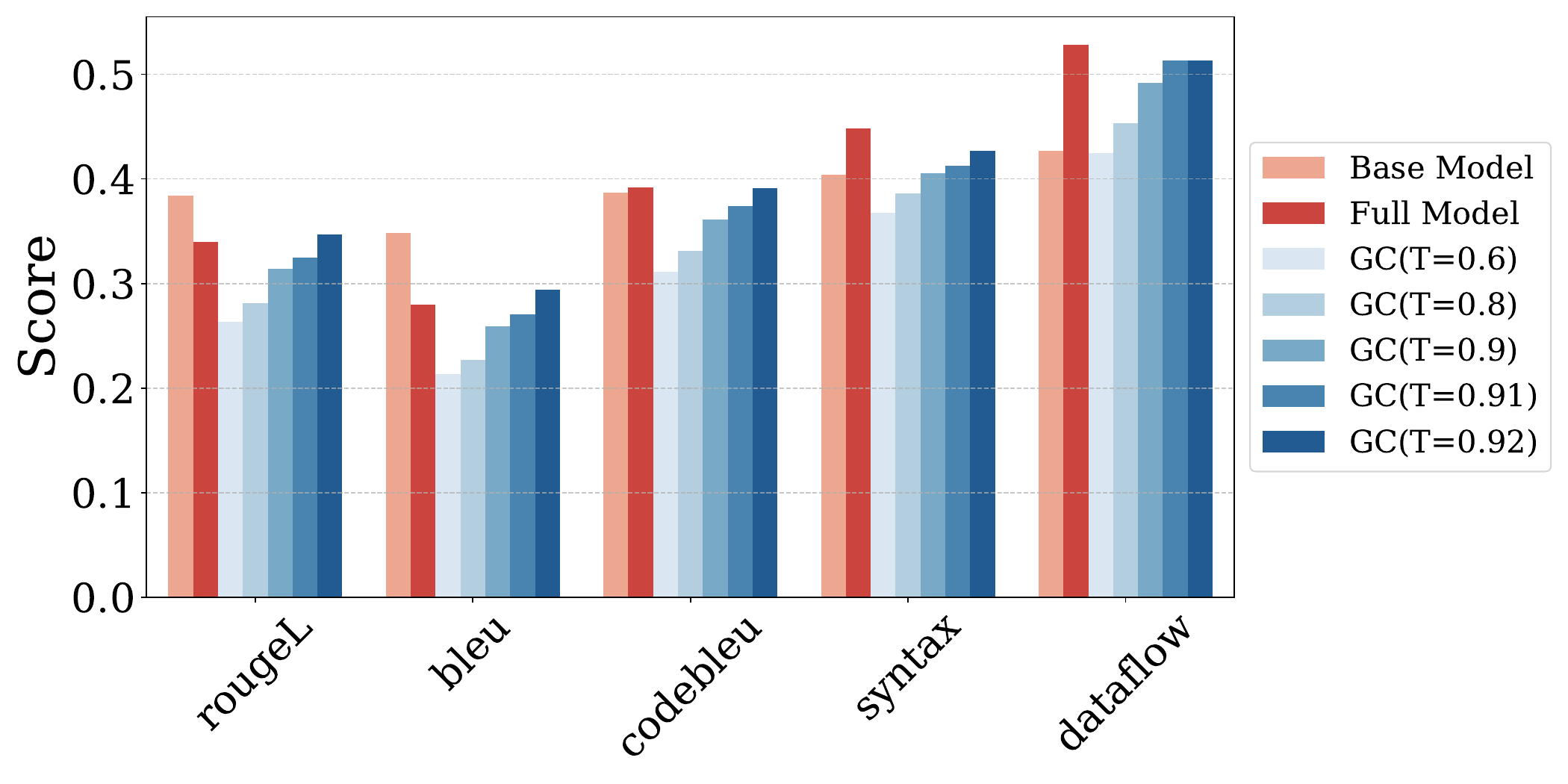}
\subcaption{Score metrics (RougeL, CodeBleu, Bleu)}
\label{llama_java_corpus_metrics:metrics_0.2}
\end{subfigure}
\begin{subfigure}[b]{0.95\columnwidth}
\centering
\includegraphics[width=\textwidth]{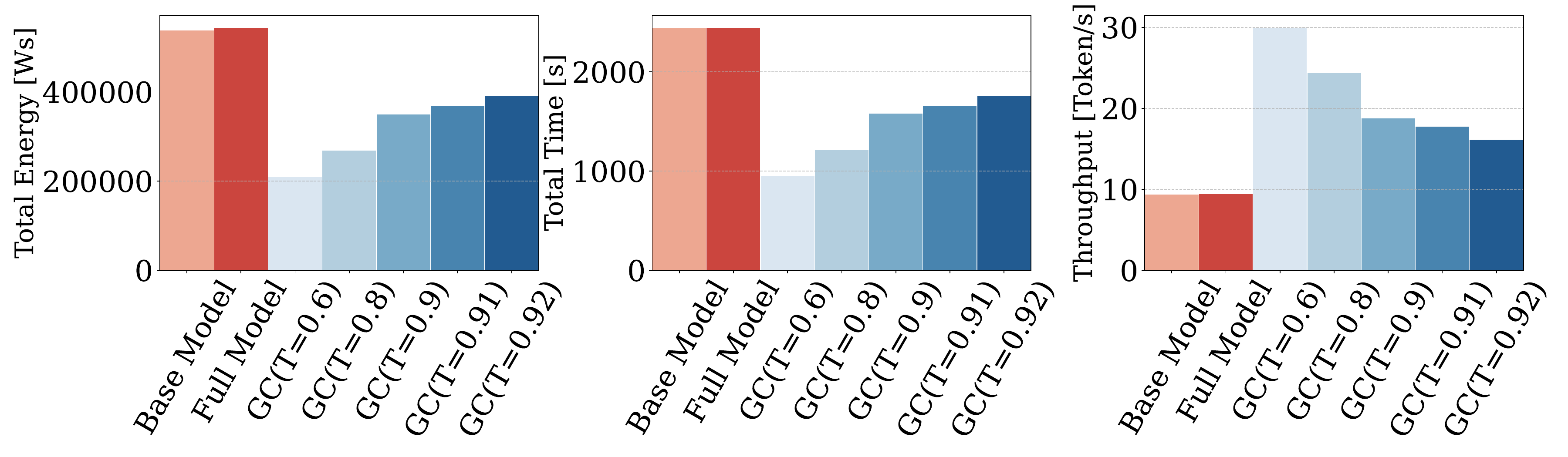}
\subcaption{Resource Consumption metrics (Energy, Time, Throughput)}
\label{opt_java_corpus_metrics:resources_0.2}

\end{subfigure}

\caption{Results of OPT on JavaCorpus, with different GC agent thresholds $T$ and baselines. }
\label{experiment:opt_java_corpus_metrics_0.2}
\end{figure}

Overall, our findings for OPT are largely comparable to those discussed for Llama. Figure~\ref{experiment:opt_java_corpus_metrics_0.2} presents the results for OPT. Here as well, we can achieve performance similar to the full-layer, e.g., 0.31 vs 0.39 CodeBLEU on least aggressive setting and 0.39 vs 0.392 on most aggressive, while saving at least 28\% of energy.

\subsubsection{Performance Evaluation on PY150 Dataset}
\begin{figure}[h]
\centering

\begin{subfigure}[b]{0.9\columnwidth}
\centering
\includegraphics[width=\textwidth]{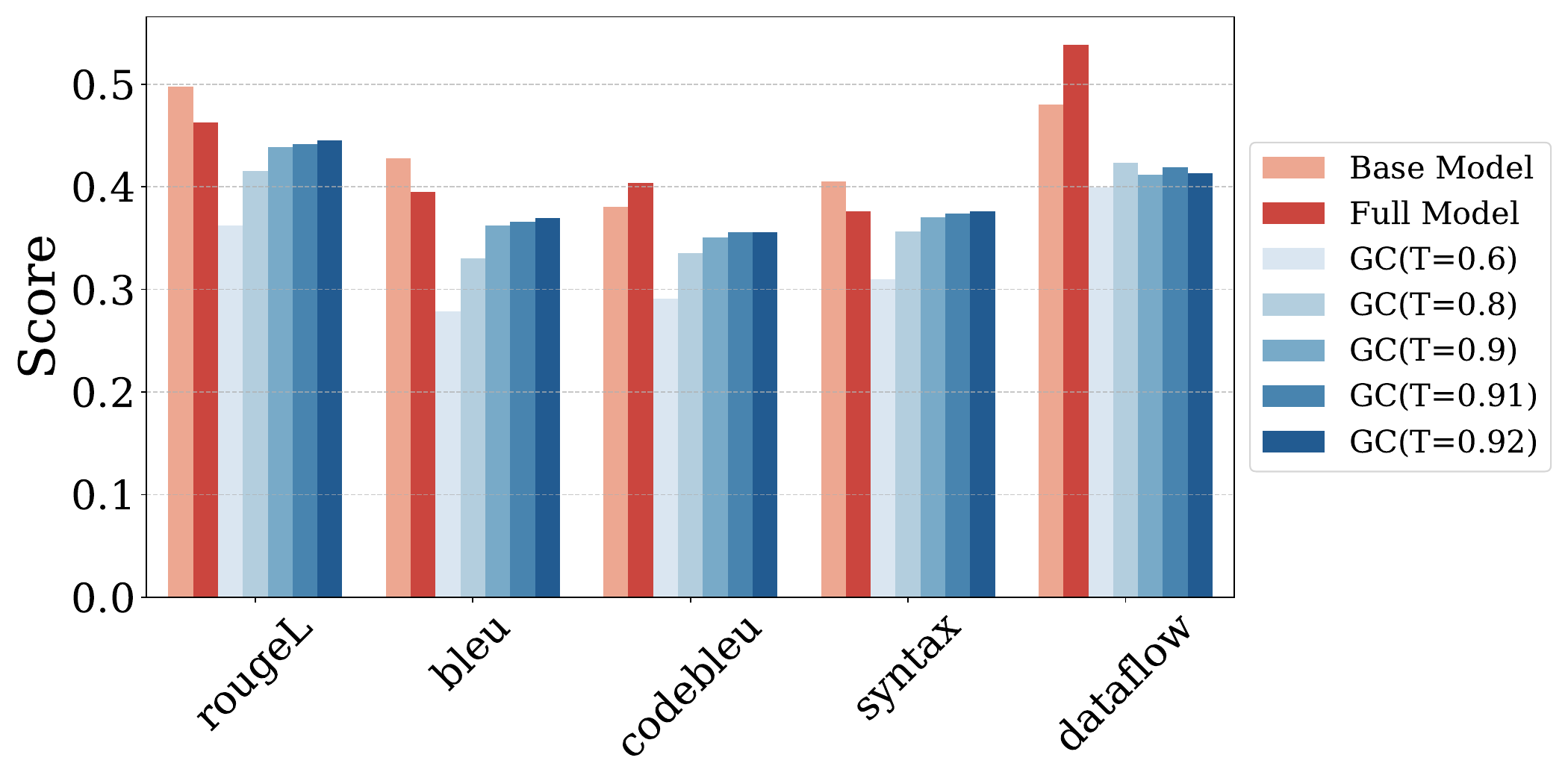}
\subcaption{Score metrics (RougeL, CodeBleu, Bleu)}
\end{subfigure}
\begin{subfigure}[b]{0.95\columnwidth}
\centering
\includegraphics[width=\textwidth]{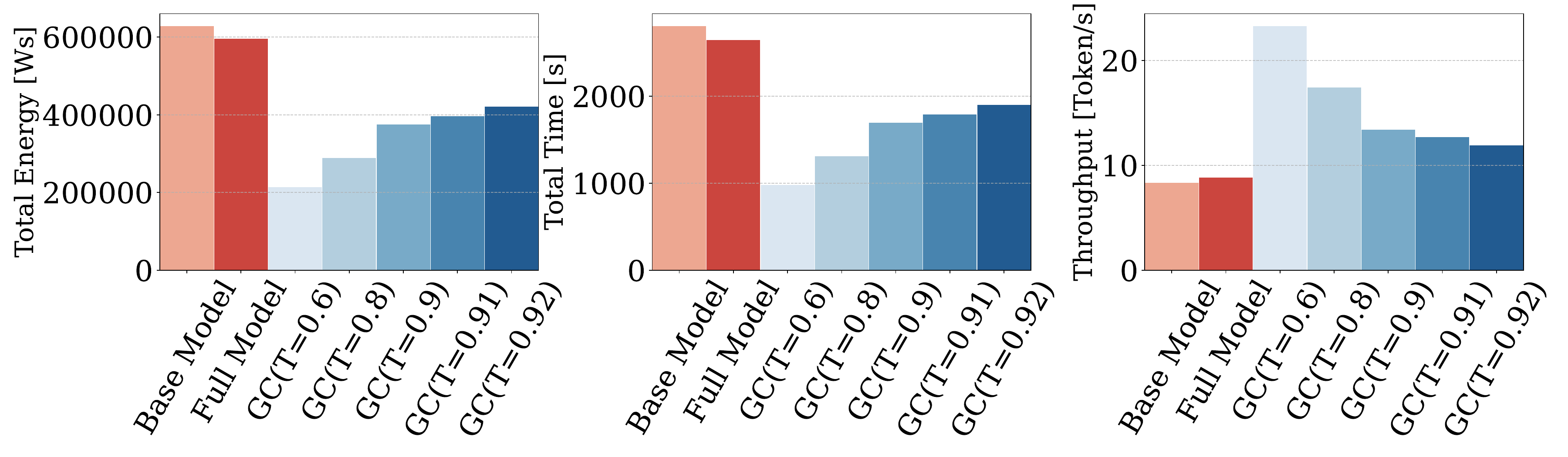}
\subcaption{Resource Consumption metrics (Energy, Time, Throughput)}

\end{subfigure}

\caption{Results of Llama on PY150,   with different GC agent thresholds $T$ and baselines. }
\label{experiment:llama_python}
\end{figure}

\begin{figure}[h]
\centering

\begin{subfigure}[b]{0.9\columnwidth}
\centering
\includegraphics[width=\textwidth]{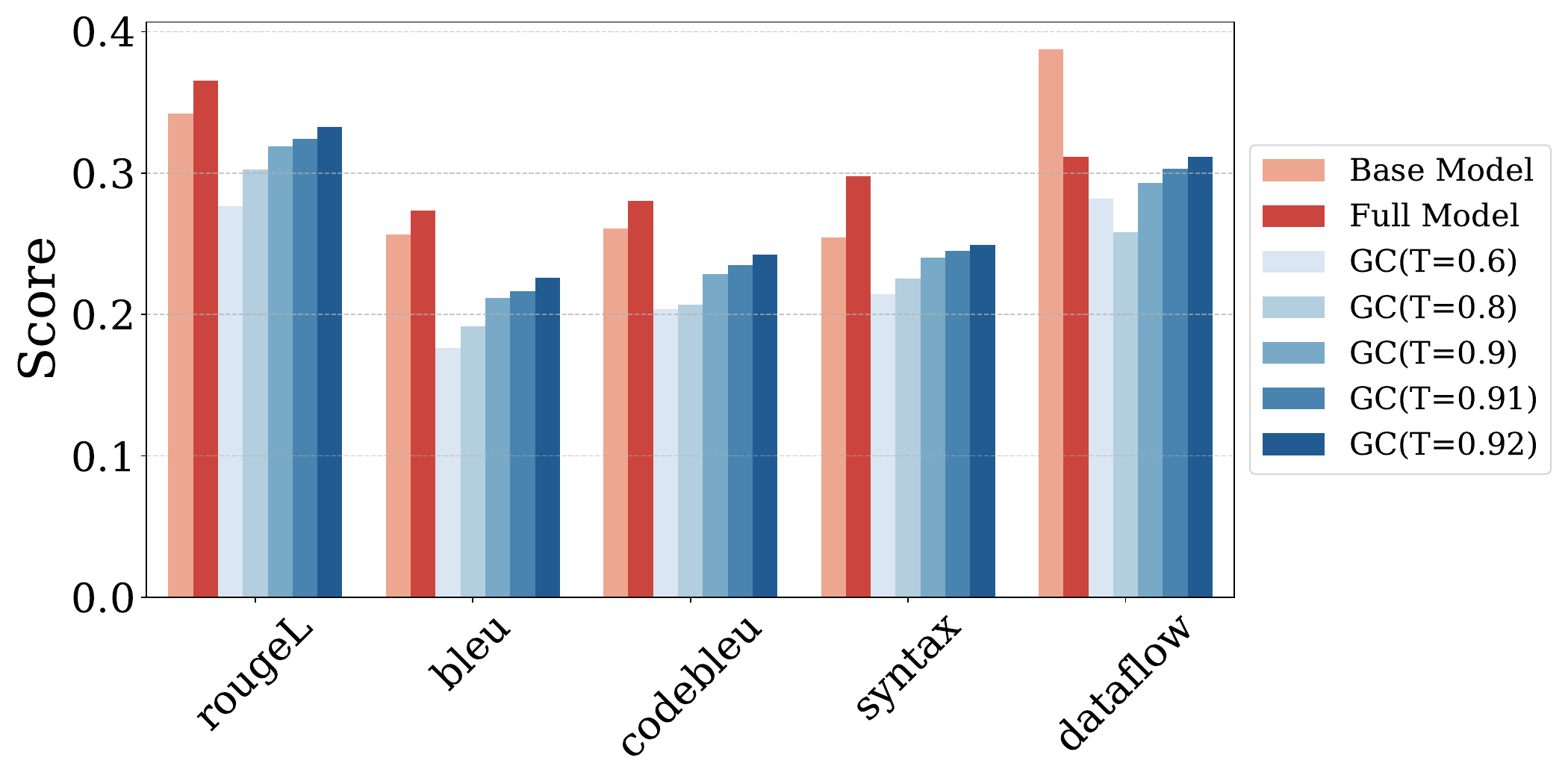}
\subcaption{Score metrics (RougeL, CodeBleu, Bleu)}
\end{subfigure}
\begin{subfigure}[b]{0.95\columnwidth}
\centering
\includegraphics[width=\textwidth]{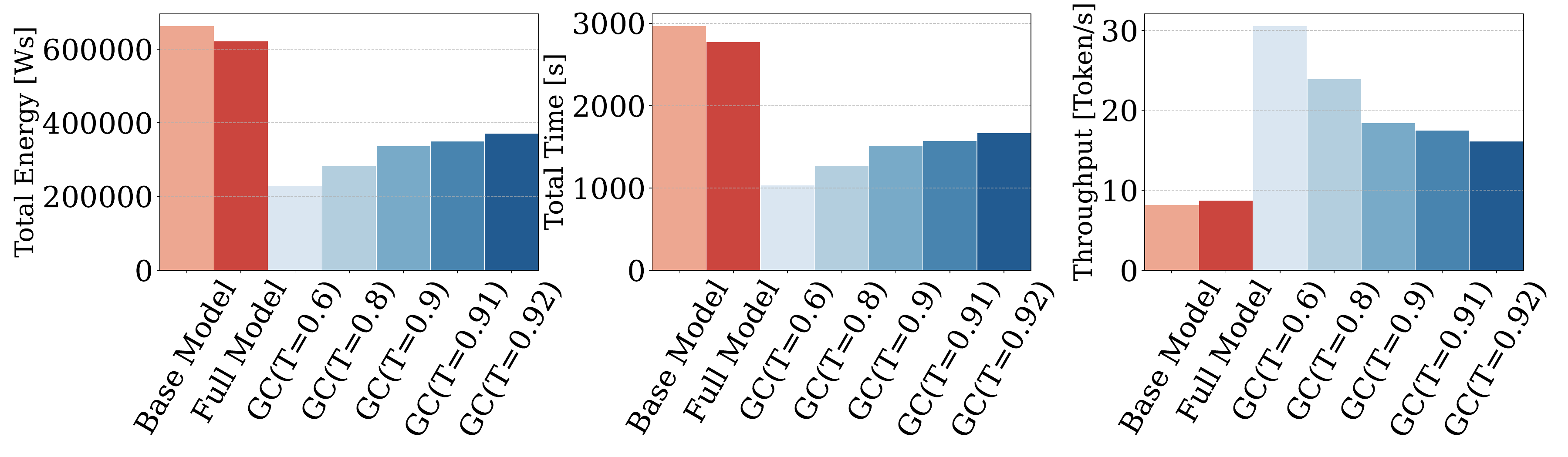}
\subcaption{Resource Consumption metrics (Energy, Time, Throughput)}

\end{subfigure}

\caption{Results of OPT on PY150, with different GC agent thresholds $T$ and baselines. }
\label{experiment:opt_python}

\end{figure}

Figure~\ref{experiment:llama_python} presents the results for Llama, and Figure~\ref{experiment:opt_python} shows the results for OPT on the PY150 dataset. Overall, the findings align with those observed for the JavaCorpus dataset. For instance, Llama achieves scores such as 0.46 vs. 0.44 for RougeL and 0.4 vs. 0.36 for CodeBLEU in the "most aggressive" setting ($T=0.92$) of GC approach, while saving approximately 29\% in energy costs compared to the model with all layers.

However, unlike the observations on the JavaCorpus dataset, we notice almost no variation between higher thresholds for the agent. This suggests that the performance is quite saturated for $T=0.9$ to $T=0.92$, and further adjustments in thresholds are unlikely to yield significant accuracy improvements.

This effect is less pronounced in the experiment with OPT, as shown in Figure~\ref{experiment:opt_python}. While the GC model enhanced with early exit achieves reasonable results overall, its performance is slightly worse compared to Llama. For instance, the full model achieves a RougeL score of 0.365, whereas the GC reaches only 0.33. Additionally, the resource-related metrics reveal that, on the Python dataset, the OPT-enhanced GC  exploits early exits more heavily than GC-Llama, which is evident on higher $T$ values in energy and time metrics (i.e., $\sim$40\% savings on OPT, $\sim$30\% on Llama).

In summary, our GC solution, enhanced with dynamic early exit through RL agent, is capable of achieving a significant reduction in energy consumption and latency metrics, with a slight trade-off with accuracy, across both model and datasets.

\subsection{Sensitivity Analysis of Context Length}
\begin{figure}[h]
\centering

\begin{subfigure}[b]{0.9\columnwidth}
\centering
\includegraphics[width=\textwidth]{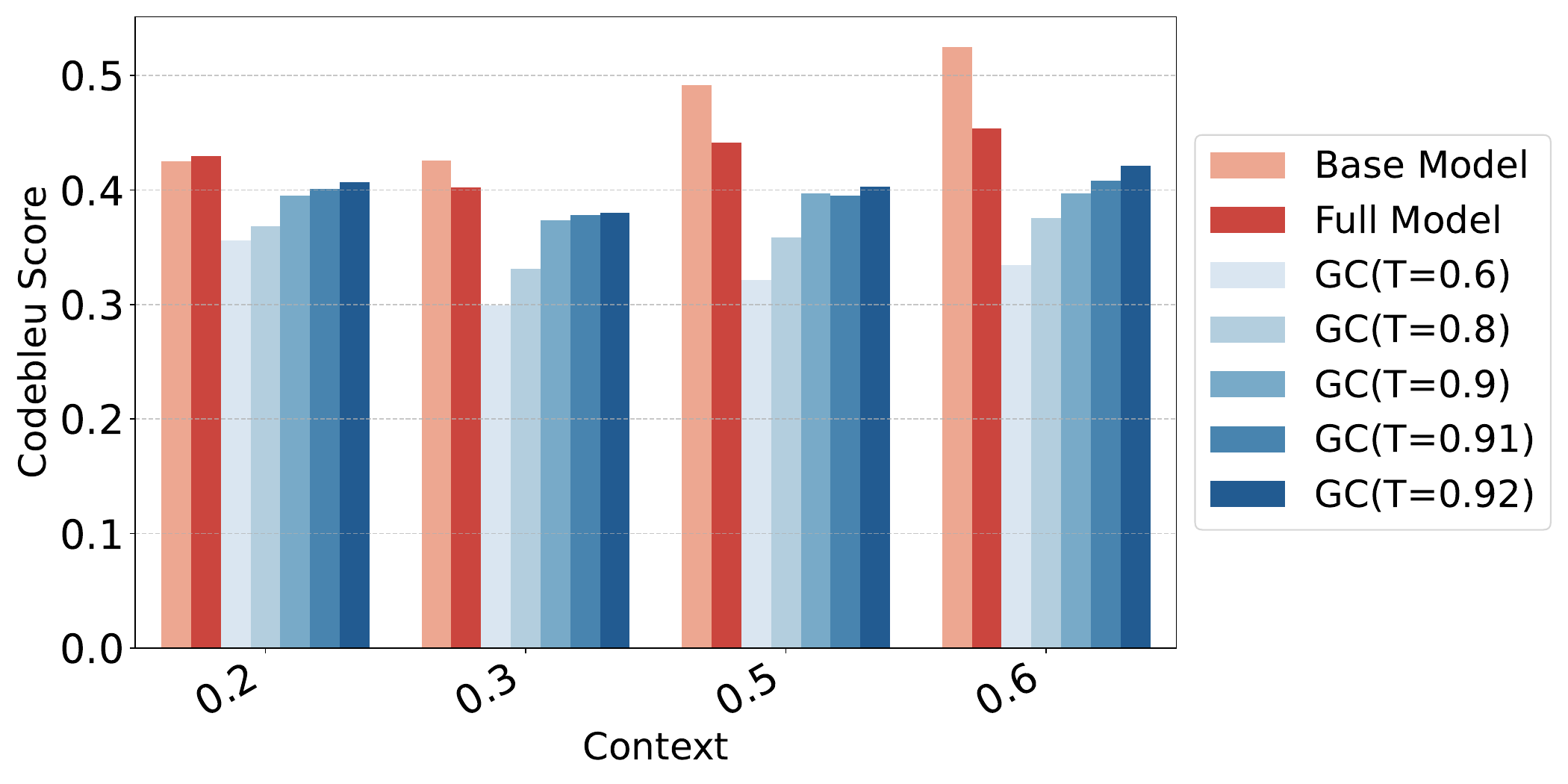}
\subcaption{CodeBLEU}
\end{subfigure}
\begin{subfigure}[b]{0.95\columnwidth}
\centering
\includegraphics[width=\textwidth]{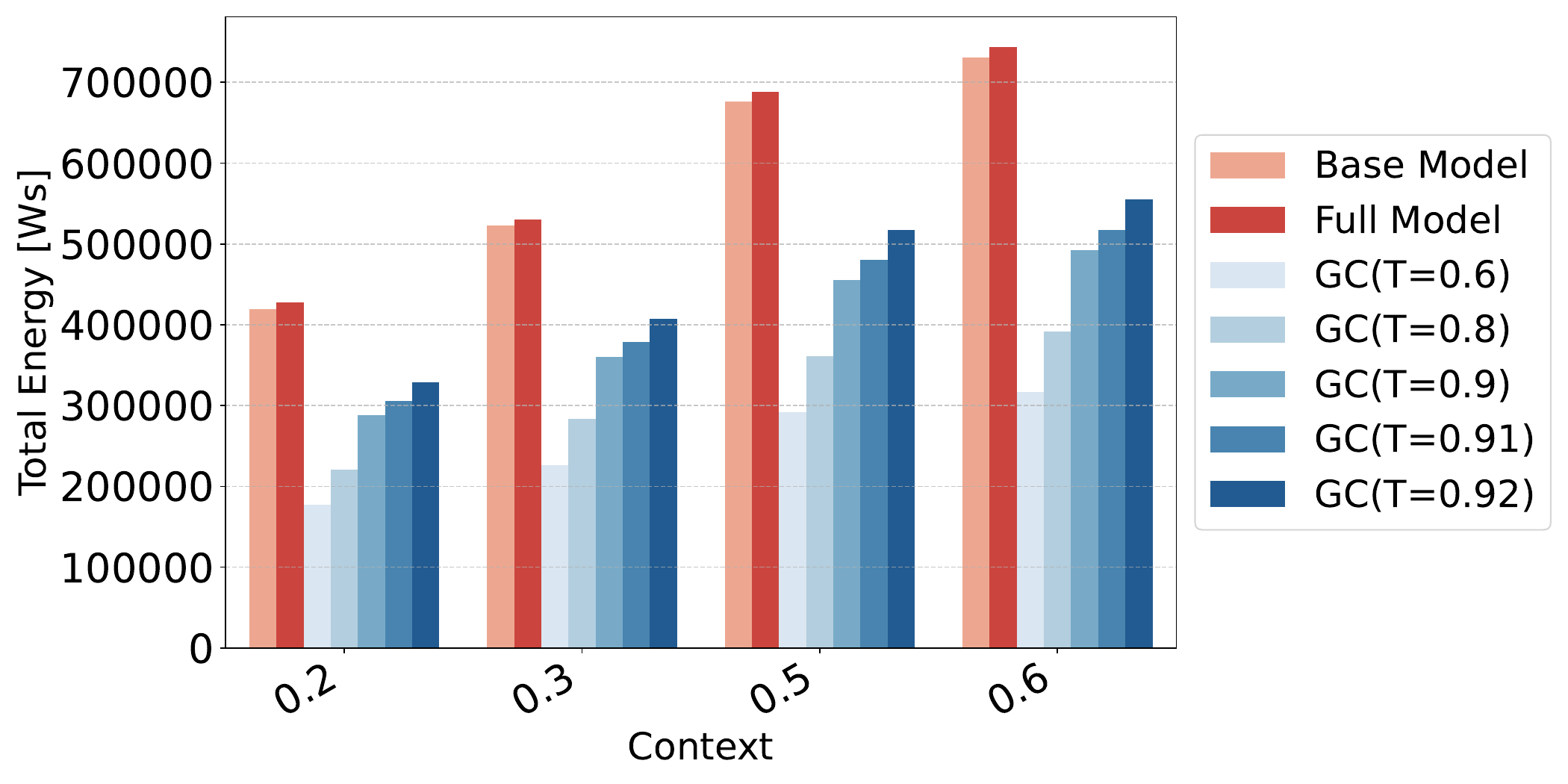}
\subcaption{Total Energy.}

\end{subfigure}

\caption{Results of Llama on JavaCorpus, with different GC agent thresholds $T$, input contexts and baselines. }
\label{experiment:llama_ctxs}

\end{figure}

In the previous subsection, we presented the context settings of the results 0.2 (that is, 20\% of the code file as input). Figure~\ref{experiment:llama_ctxs} illustrates the performance of Llama3.2 on JavaCorpus in different context lengths of 0.2, 0.3, 0.5, and 0.6, while varying the agent thresholds as described earlier. 
The results are shown for CodeBLEU and energy, as these metrics exhibit overall behavior similar to other indicators.

The findings demonstrate that GREEN-CODE overall performs similarly in all context settings. However, in some higher context settings, the difference to the full model is larger than in lower context settings. 
For example, in the context setting 0.2, CodeBLEU is 17\% lower at $T=0.6$, but only 5.28\% lower at $T=0.92$. In contrast, in the context of 0.5, these values increase to 27.2\% and 8.9\%, respectively, almost double the loss observed in the context of the lower context. Similar patterns are evident in the other context settings.
Although there is a small variation in performance across different input sizes, its behavior in terms of energy savings remains similar. Consequently, we reported primary results for context setting 0.2.

\subsection{Analyzing KV Cache Impact }
\begin{figure}[h]
\centering

\begin{subfigure}[b]{0.63\columnwidth}
\centering
\includegraphics[width=\textwidth]{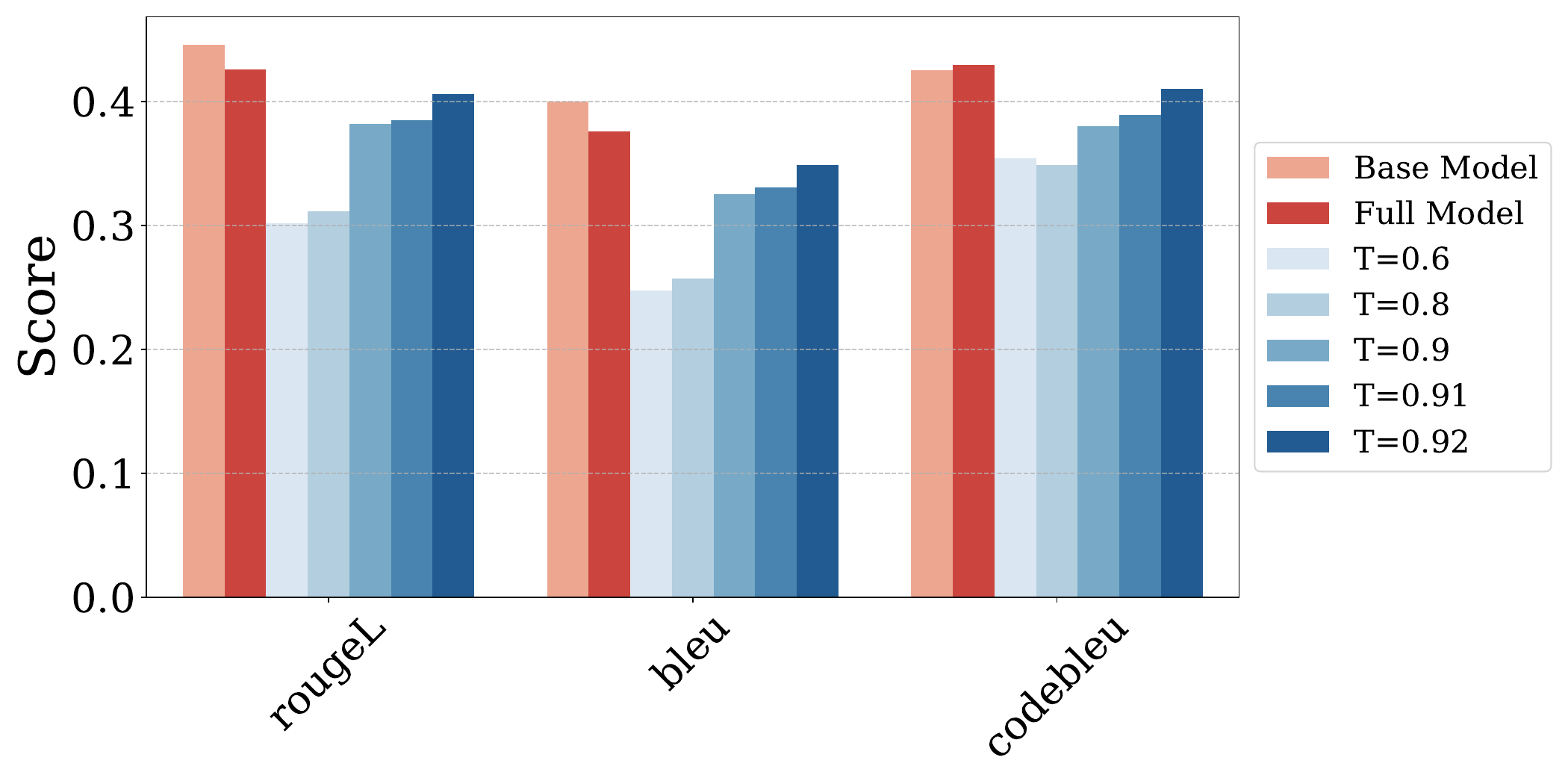}
\subcaption{Score metrics}
\end{subfigure}
\begin{subfigure}[b]{0.35\columnwidth}
\centering
\includegraphics[width=\textwidth]{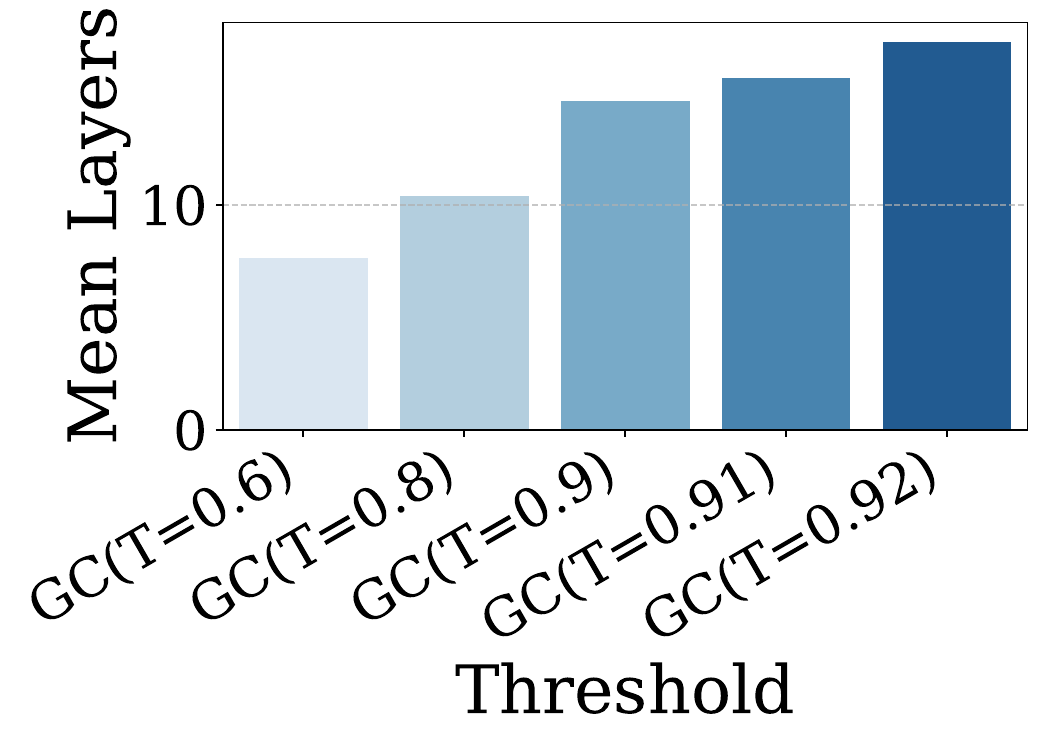}
\subcaption{Avg. \#Layers.}

\end{subfigure}

\caption{Results of Llama on JavaCorpus with KV caching. }
\label{experiment:llama_kv}

\end{figure}

The experiments described above were conducted without KV caching due to the native incompatibility of KV caching with early exits. This incompatibility arises because caches may be unavailable when a deeper exit follows a shallower one. Although this aspect was not our primary focus, we implemented a basic KV cache propagation method, as outlined in previous studies~\cite{Schuster2022}, to evaluate the impact of KV caching. Figure~\ref{experiment:llama_kv} presents the results, including various accuracy-related metrics and the average number of layers utilized per experiment. The findings indicate that our approach achieves accuracy comparable to other reported methods. Optimizing KV caching during early exits, such as through parallel cache computation, remains a promising avenue for future research.

\subsection{Overhead Analysis}
GREEN-CODE introduces computational overhead in two parts, (i)  forward pass through the RL network and (ii) softmax computation to decide the agent's action for a given threshold.

We conducted an isolated experiment measuring energy and latency to eliminate inaccuracies caused by nested measurements. Table~\ref{exper:overhead} presents the relative energy and time overhead of both OPT and Llama models for two variations each: (i) a model  with EE enabled and (ii) full model without exits.

The results show that the additional energy required for the RL agent is reasonable compared to the overall energy consumption of a full model. Moreover, the overhead slightly increases as the RL agent thresholds rise. This is because higher thresholds result in more "continue" actions being chosen, leading to more additional "exit-checks" from RL agent. Furthermore, we can see that the overhead on OPT is overall higher than on Llama, which is expected due to the additional layers. Overall, the overheads are always below $1/5^{th}$ of the total runtime when using the EE-enhanced model, indicating the practical usefulness of GREEN-CODE's solution approach.  Furthermore,  the energy overhead is slightly higher than the latency overhead, which could be attributed to  factors, such as forward passes being fast yet energy-intensive.

\begin{table}[!t] 
\centering
\scriptsize
    \resizebox{\columnwidth}{!}{%
    \begin{tabular}{lllll} 
    \toprule
    \textbf{$T$} & \textbf{OPT} & \textbf{OPT to full model}  & \textbf{Llama} & \textbf{Llama to full model} \\
  \midrule

    0.6 & 17.68\%/11.19\% & 4.38\%/4.85\% & 11.39\%/7.47\% & 3.32\%/3.07\%\\
    0.8 & 18.14\%/11.29\% & 5.55\%/6.04\%& 14.79\%/9.53\%&5.35\%/5.41\%\\
    0.9 & 18.78\%/11.7\% & 6.78\%/7.53\% & 16.39\%/10.63\% & 7.33\%/7.94\%\\
    0.91 & 19.06\%/11.80\%  & 7.12\%/7.9\% & 17.16\%/10.88\%  & 8.05\%/8.61\%\\
    0.92 & 19.43\%/11.98\%& 7.66\%/8.53\%& 17.94\&/11.12\% & 8.83\%/9.37\%\\
 \bottomrule
    \end{tabular}} 
    \caption{Relative overhead on PY150 with different thresholds. Reported as pairs of Overhead-Energy/Overhead-Time.}
    \label{exper:overhead}
\end{table}

\label{sec:results}
\section{Related Work}

Pruning techniques, such as LLM-pruner~\cite{ma2023}, LaCo~\cite{yang2024} or SoBP~\cite{wei2024}  aim to reduce the model size by decreasing the number of layers or weight blocks, thus creating lightweight models that require smaller resource consumption. Similarly, quantization techniques~\cite{Dettmers2024,Chee2024,Kuzmin2022,Liu2021} aim to reduce the model size by using lower precision weights, e.g., converting floating point numbers to small integer representations. Another method is knowledge distillation (KD)~\cite{jiao2020,  li2024, timiryasov2023}, where the original model teaches a smaller student model its outputs. By learning the predictions and complexities of a teacher, the student aims to mimic the behavior of the larger model and is more efficient as it is smaller. However, all these model preprocessing techniques reduce model performance irreversibly and do not adapt to dynamic contexts.

The studies most relevant to our work are early exiting methods.
The work in ~\cite{Schuster2022} introduces early exiting on top of T5 models with confidence scores or using classifiers to decide when to early exit. 
BERxIT~\cite{xin2021} improves on this by introducing learn-to-exit modules. The LayerSkip method~\cite{elhoushi2024} uses dropout during training to enable early exiting and employs self-speculative decoding, where early exited results are verified and corrected by the remaining layers of the model. The most relevant work to us is Sun et al.~\cite{Sun2024}, which performs early exiting on code generation. Their method introduces a classifier-based exiting approach with multiple LM heads and additional stopping of inference if a prediction seems non-promising. Similar to our work, ConsistentEE~\cite{Zeng2024} uses reinforcement learning for exiting, introducing policies and LM heads at every layer where exiting is possible. However, both these works introduce additional overhead with multiple LM heads. Furthermore, the LITE framework~\cite{varshney2024} introduces aggregation from losses of intermediate layers to enable EE with the original LM head, but only uses confidence-based exiting on instruction tuning tasks. Finally, other EE techniques include up-casting hidden representations to skip layers~\cite{yom-din2024} or learning instance difficulty~\cite{TSun2022}. Our work distinguishes itself by providing a framework to early exit dynamically at arbitrary layers by adapting to changing requirements without introducing overheads of multiple LM heads.
\label{sec:related_work}
\section{Conclusions}

Efficient early exiting strategies can significantly reduce energy consumption and computational latency in LLM code generation tasks. However, identifying the right exit layer without degrading model accuracy is extremely challenging due to the complexities of model architectures and input distributions. To address this, we present GREEN-CODE, a framework that dynamically selects early exit points for given input tokens of a code file. GREEN-CODE employs a customized fine-tuning method with weighted aggregate loss to adapt base pre-trained LLMs to exit at intermediate layers with a single LM head. Furthermore, we frame the early exiting problem as an RL decision problem and train the RL agent, which dynamically chooses the right exit layer during code completion tasks. Our experiment results demonstrated that GREEN-CODE achieves comparable accuracy to the non-fine-tuned base model using all layers, with significant reductions in energy consumption and latency, demonstrating the feasibility of such techniques in real-world usage.

In the future, we plan to extend this framework to incorporate larger models for different LLM tasks. We also intend to consider complex multi-agent-based software engineering workflows, where long-running inference workflows benefit more with an early exiting strategy.
\label{sec:conclusions}

\section*{Acknowlegments}

The research presented in this paper has been partially
funded through the following projects: High-Performance Integrated
Quantum Computing (HPQC), Austrian Research Promotion
Agency (FFG) Nr. 897481; Transprecise Edge Computing (Triton), Austrian Science Fund (FWF), DOI: 10.55776/P36870;
and Trustworthy and Sustainable Code Offloading (Themis), FWF, DOI: 10.55776/PAT1668223.

\bibliographystyle{IEEEtran}
\bibliography{refs}
\vspace{12pt}
\end{document}